\documentclass[aps,prb,floatfix,amsmath,amssymb,preprint,eqsecnum,nofootinbib]{revtex4}
\usepackage{graphicx}
\usepackage{dcolumn}
\usepackage{bm}
\usepackage{amsmath, amsfonts, amssymb}
\usepackage{pstricks}
\usepackage{amsxtra}
\usepackage{amsthm}

\usepackage{soul} 

\newcommand{\bdag}{\hat{b}^\dag}

\renewcommand{\ddag}{\hat{d}^\dag}

\newcommand{\pdag}{\hat{p}^\dag}
\newcommand{\hdag}{\hat{h}^\dag}
\newcommand{\teff} {t_{\rm eff}}
\newcommand{\bra}[1]{\left\langle#1\right|}
\newcommand{\ket}[1]{\left|#1\right\rangle}
\newcommand{\braket}[2]{\left\langle#1|#2\right\rangle}
\newcommand{\abs}[1]{\left|#1\right|}
\newcommand{\ave}[1]{\langle#1\rangle}

\newcommand{\eqnref}[1]{Eq.~(\ref{#1})} 
\newcommand{\fref}[1]{Fig.~\ref{#1}}    
\newcommand{\secref}[1]{Section~\ref{#1}}

\newcommand{\hc}{\mathrm{H.c.}}

\newcommand{\beq}{\begin{equation}}
\newcommand{\eeq}{\end{equation}}
\newcommand{\ba}{\begin{array}{ccc}}
\newcommand{\ea}{\end{array}}

\def\bea{\begin{eqnarray}}
\def\eea{\end{eqnarray}}

\usepackage{setspace} 
\begin{document}
\title{Correlated phases of bosons in tilted, frustrated lattices}
\author{Susanne Pielawa}
\affiliation{Department of Physics, Harvard University, Cambridge MA
02138}
\author{Takuya Kitagawa}
\affiliation{Department of Physics, Harvard University, Cambridge MA
02138}
\author{Erez Berg}
\affiliation{Department of Physics, Harvard University, Cambridge MA
02138}
\author{Subir Sachdev}
\affiliation{Department of Physics, Harvard University, Cambridge MA
02138}

\date{\today \\
\vspace{1.6in}}
\begin{abstract}

We study the `tilting' of Mott insulators of bosons into metastable states.
%
%
These are described by Hamiltonians acting on resonant subspaces, and have rich possibilities for correlated phases with non-trivial entanglement of pseudospin degrees of freedom encoded in the boson density. 
We extend a previous study (Phys. Rev. B {\bf 66}, 075128 (2002)) of cubic lattices to a
variety of lattices and  tilt directions in 2 dimensions: square, decorated square, triangular, and kagome. 
For certain configurations three-body interactions are necessary to ensure that the energy of the effective resonant subspace is bounded from below. 
We find quantum phases with Ising density wave order, with superfluidity transverse
to the tilt direction, 
and a quantum liquid
state with no broken symmetry. 
The existence of the quantum liquids state is shown by an exact solution for a particular correlated boson model. 
We also find cases for which the resonant subspace is described by effective quantum dimer models.


\end{abstract}

\maketitle

\section{Introduction}
\label{sec:intro}
The observation of the superfluid-insulator quantum phase transition of ultracold
bosonic atoms in an optical lattice \cite{bloch} has launched a wide effort to engineering
other correlated quantum states of trapped atoms. Much of the effort has focused on entangling the spin quantum
number of the atoms between different lattice sites: superexchange interactions between neighboring lattice
sites have been observed \cite{super}, but longer range spin correlations have not yet been achieved.

Here we focus on a proposal \cite{SachdevMI,ss2} to obtain quantum
correlated phases in a pseudospin degree of freedom which measures
changes in boson number across the links of a tilted lattice
\cite{Pachos, twoband}. The interactions which entangle the pseudospins are
not suppressed by factors of $t/U$ (where $t$ is the tunneling
between lattice sites, and $U$ is the local boson repulsion), and
so non-trivial entanglement is likely to be readily accessible\cite{1dIsingExp}.
This proposal takes advantage of the unique feature of ultracold
atom systems, namely, the ability to study many-body
non-equilibrium physics\cite{1dIsingExp,esslinger, kinoshita, schmiedmayer,
kitagawa}, and suggests that various intriguing phases can appear
as metastable states\cite{huber}. Moreover, the recent advent of
site-resolved imaging of ultracold atoms in optical lattices
\cite{bakrgreiner2, BakrGreiner,Sherson} opens the way to the
direct detection of the pseudospin and the investigation of its
quantum correlations.

The previous theoretical work on tilted lattices \cite{SachdevMI,ss2,ss3} focused on simple cubic lattices
in spatial dimensions $d=1,2,3$, with the tilt along one of the principal cubic axes. In $d=1$, this theory
predicted Ising density wave order for strong tilt, and this has been observed in recent experiments\cite{1dIsingExp}.
Here, we will
study a variety of other lattices and tilt directions. We will find analogs of the density wave and anisotropic superfluid phases found earlier. However we will also find new quantum liquid states, and some cases which map onto the quantum dimer model. These
states entangle the displacements of all the atoms, and so should be attractive targets of future experiments.
We will find an exact liquid ground state for a particular correlated boson model.

We begin by recalling the Hamiltonian of a tilted Mott insulator.
It is described by the generalized bosonic Hubbard model\cite{Jaksch} with an additional potential gradient along a certain direction:
$H=H_{\rm kin}+H_{\rm U}+H_{\rm tilt}$:
\begin{subequations}
\begin{eqnarray}
H_{\rm kin} &=& - t \sum_{<ij>} \left( \bdag_i \hat b_j + \bdag_j \hat b_i \right) \\
H_{\rm U}&=&\frac U 2 \sum_i \hat n_i (\hat n_i-1) + \frac {U_3} {6} \sum_i n_i (n_i-1)(n_i-2)
+\dots \\
\label{eq:repulsion}
H_{\rm tilt}&=&- E \sum_i {\bf e} \cdot {\bf r}_i \, \hat n_i .
\label{hubbard}
\end{eqnarray}
\end{subequations}
Here $\hat b_i$ are canonical boson operators on lattice sites $i$ at spatial co-ordinate
${\bf r}_i$, and $\hat n_i \equiv \bdag_i \hat b_i$.
The first term in $H_{\rm U}$ describes two body interactions. The
second term is an effective three body interaction, generated by
virtual processes involving higher bands \cite{multibody,
threebody}. Such a term is present in ultracold atomic systems, as
has recently been measured\cite{multibody, threebody, Waseem}. As
we will show, the presence of this term dramatically changes the
physics of tilted two dimensional lattices, therefore we include
it in our model.
The potential gradient is $E$, and the fixed vector
${\bf e}$ is normalized so that the smallest change in potential energy between neighboring lattice sites
has magnitude $E$.
%
%
We will assume a filling of one atom per site in the parent Mott insulator in most of the following discussion.

We are interested in the dynamics of the resonant subspace that appears when the lattice is tilted
so that $E$ becomes of order $U$. Specifically, we will work in the limit
\begin{equation}
\abs{\Delta}, t \ll E,U , \label{inequality}
\end{equation}
where we define
\begin{equation}
\Delta=U-E .
\end{equation}
We also define our tuning parameter
\begin{equation}
\lambda \equiv \frac {\Delta}{t},
\end{equation}
and allow $\lambda$ to range over all real values. Thus we include
processes which carry energy denominators of order $U-E$ to all
orders, but neglect processes which have energy denominators of
order $U$ or $E$.

\begin{figure}
\begin{minipage}{.6\textwidth}
\includegraphics[width=.6\textwidth]{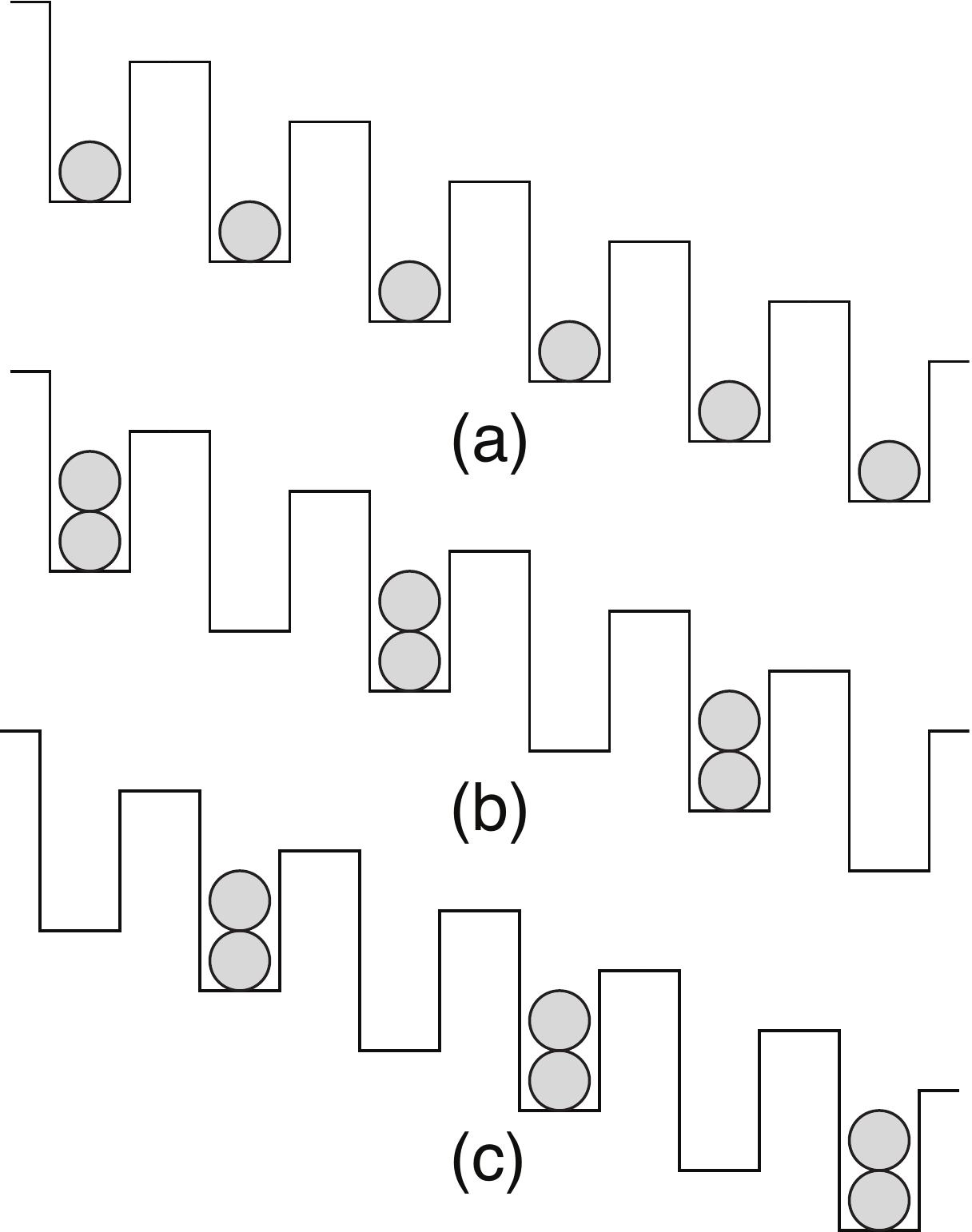}
\end{minipage}
\caption{(a) Parent Mott insulator, which is the ground state for $\lambda \rightarrow \infty$.
(b,c) Ground states with Ising density wave order for $\lambda \rightarrow - \infty$.}
\label{isingd1}
\end{figure}

Let us now review the properties \cite{SachdevMI} of a tilted chain of sites in $d=1$;
see \fref{isingd1}.
As explained in Ref.~\onlinecite{SachdevMI}, the limit \eqnref{inequality}
defines a strongly interacting many-body
problem in the resonant subspace, which can be described by an effective Hamiltonian.
Unlike the underlying Hubbard model in Eq.~(\ref{hubbard}), the energy of this effective Hamiltonian
is bounded from below, and it makes sense to study its stable ground state and low-lying excitations by the
traditional methods of equilibrium many body theory. This will help us describe the dynamics of experimental
studies in the metastable resonant subspace of the full Hubbard model in Eq.~(\ref{hubbard})
defined by Eq.~(\ref{inequality}).
It was found that the resonant subspace was described by a quantum Ising model for an Ising pseudospin residing
on the links of the chain: an up spin represented the transfer of an atom across the link
(or the creation of a dipole particle-hole pair), while the down spin had no change from the parent
Mott insulator. In other words, in a parent Mott insulator with occupancy $n_{0}$ per site, a down spin was a pair of sites
with occupancy $(n_{0},n_{0})$, while an up spin was $(n_{0}-1,n_{0}+1)$. The Ising model had both a longitudinal and transverse external field, representing the energy of the dipole
and the tunneling amplitude of the atom respectively. However, two nearest neighbor dipoles cannot
be created simultaneously, and this implied the presence of infinitely strong nearest-neighbor {\em antiferromagnetic\/} exchange interactions.
It is this infinitely strong exchange which allows easy access to strong-correlation physics
in such tilted lattices\cite{1dIsingExp}.
The phase diagram of the Ising antiferromagnet with both longitudinal and transverse fields has also been studied by
others, \cite{novotny, Ovchinnikov} with different physical motivations.
For $\lambda \rightarrow \infty$, the Ising model had a paramagnetic ground state,
corresponding to configurations close to the parent Mott insulator. In contrast for $\lambda \rightarrow - \infty$,
the Ising model had antiferromagnetic long-range order, corresponding to a maximum density of particle-hole
pair excitations about the Mott insulator. There are two possible ways of arranging these excitations,
depending upon whether the particles reside on the odd or even sites, and this ordering is linked to the antiferromagnetic
order of the Ising model. In between these limit cases, a quantum phase transition was found \cite{SachdevMI} at
an intermediate value of $\lambda$: this transition was in the universality class of the quantum Ising {\em ferromagnet}
in a transverse field and zero longitudinal field \cite{novotny, Ovchinnikov,SachdevMI}.


Next, let us review \cite{SachdevMI} cubic lattices in dimensions $d>1$, with a tilt along a principal lattice axis,
with $\abs{U_3}\gg t$.
Then we have to distinguish
the physics parallel and transverse to the tilt direction. Parallel to the tilt direction, the physics is similar to $d=1$:
there is an Ising density wave order associated with dipolar particle-hole pairs,
which turns on as $\lambda$ decreases. In contrast, transverse to the tilt direction, the quantum motion of individual
particles and holes is allowed, not just of dipole bound states. Such motion raises the possibility of
Bose condensation of particles/holes and of the appearance of superfluidity. However, the single particle
or hole motion is constrained to be purely transverse. Consequently there are an infinite number of conservation
laws, one for each lattice layer orthogonal to the tilt direction: the total number of particles in any layer is constrained
to equal the total number of holes in the preceding layer. Because of these conservation laws, the superfluidity
is not global, but restricted to each transverse layer separately: there is no Josephson tunneling term which
links the superfluid order parameter of adjacent layers. Of course, because of the translational symmetry
in the effective Hamiltonian parallel to the tilt, the superfluidity appears simultaneously in all the layers.
In this superfluid state, atomic motion is insulating parallel to the tilt, and superfluid transverse to the tilt.
Such transverse-superfluid states also appear as $\lambda$ decreases,
and can co-exist with the Ising order parallel to the tilt.
The previous work
\cite{SachdevMI} did not note the $U_{3}$ term, even though its presence is required for the applicability of the $d>1$ dimensional results therein.

In the present paper we will examine several tilted two dimensional lattices. We find that the physics not only depends
on the lattice geometry and tilt direction, but also on the magnitude of $U_{3}$ compared to other energy scales of the systems. 
This $U_3$ term is needed to guarantee the stability of some
lattices, as for example the cubic lattices reviewed above. The
reason is the following. Let us assume a parent Mott insulator
with filling $n_0=1$. Processes of the kind
$(1,1)\rightarrow(0,2)$ cost repulsion energy $U$, and gain
tilt-energy $E$ (if the two participating sites differ in
potential energy $E$). These processes are tuned to be resonant.
Depending on the lattice structure, it may then happen that two
sites differing in potential $E$ are each occupied by two bosons.
The process $(2,2)\rightarrow(1,3)$  costs repulsion energy
$U+U_3$, and gains potential energy $E$. In the case $\abs{U_3}\gg
t$ this process is off-resonant. We can then identify a resonant
subspace which focuses on the $(1,1)\rightarrow(0,2)$ processes
only. This was implicitly assumed in Ref.~\onlinecite{SachdevMI}
without discussion of the crucial role played by $U_{3}$, and will
be assumed in our
Sections~\ref{sec:square}-\ref{decorated_lattices}. In such cases,
our effective Hamiltonian\cite{comment-symmetries} method applies,
the energy of the resonant subspace is bounded from below, and we
can use equilibrium methods to study the phase diagrams.

On the other hand, if $\abs{U_3} \lesssim t$, we cannot neglect
processes of the type $(2,2)\rightarrow(1,3)$. In the presence of
such processes, the resonant Hilbert space is increased. This can
have dramatic consequences on the physics; in some cases, we find
that the energy in the resonant subspace becomes unbounded from
below, and the system becomes unstable.


The present paper will examine several lattice configurations of Eq.~(\ref{hubbard}) in $d=2$.
We will assume $\abs{U_3}\gg t$ in all cases, except in \secref{sec:tetris}.
We will find analogs of the Ising and transverse-superfluid phases described above in these
lattices. However, we will also find a qualitatively new phenomenon: the appearance of
quantum liquid ground states with no broken symmetries. These phases appear
in these lattices for two main reasons.
First, even for a large tilt, not all the sites can participate in the creation of
particles or holes due to the geometrical constraints present in these lattice configurations.
Such geometrical constraints generate exponentially large degenerate states
in the limit of infinite tilt. Secondly, the lattice
structure ensures that there is no free motion of unbound particles or holes in the direction
transverse to the tilt in the resonant subspace. The absence of such unbound motion prevents
transverse-superfluidity, allowing for the appearance of the quantum liquid states.

The outline of the remainder of the paper is as follows.
In \secref{sec:square} we study a square lattice with a tilt in a diagonal direction, while in \secref{sec:triangular}
we study a triangular lattice with different tilt directions. In these cases, we will find analogs of phases
studied previously, involving Ising density wave and transverse-superfluid order.
In \secref{decorated_lattices}, we will consider the kagome and certain decorated square lattices.
Here, the transverse-superfluid order is suppressed for certain tilt directions, and novel quantum liquid
ground states appear. Also, for another tilt direction on the kagome, the Ising density wave order
is suppressed, and we obtain 
effective decoupled one dimensional systems.
In \secref{sec:tetris},
we briefly discuss the $U_3=0$ case.

\section{Square lattice}
\label{sec:square}
As noted in Section~\ref{sec:intro}, this section always assumes $\abs{U_{3}} \gg t$.
We have already reviewed the results of Ref.~\onlinecite{SachdevMI} for a square lattice tilted
along a principle lattice direction. Ref.~\onlinecite{SachdevMI} also briefly considered a tilt
along a diagonal lattice direction, and we will discuss this case more completely in the present section.

\begin{figure}
\begin{center}
\includegraphics[width=3.5in]{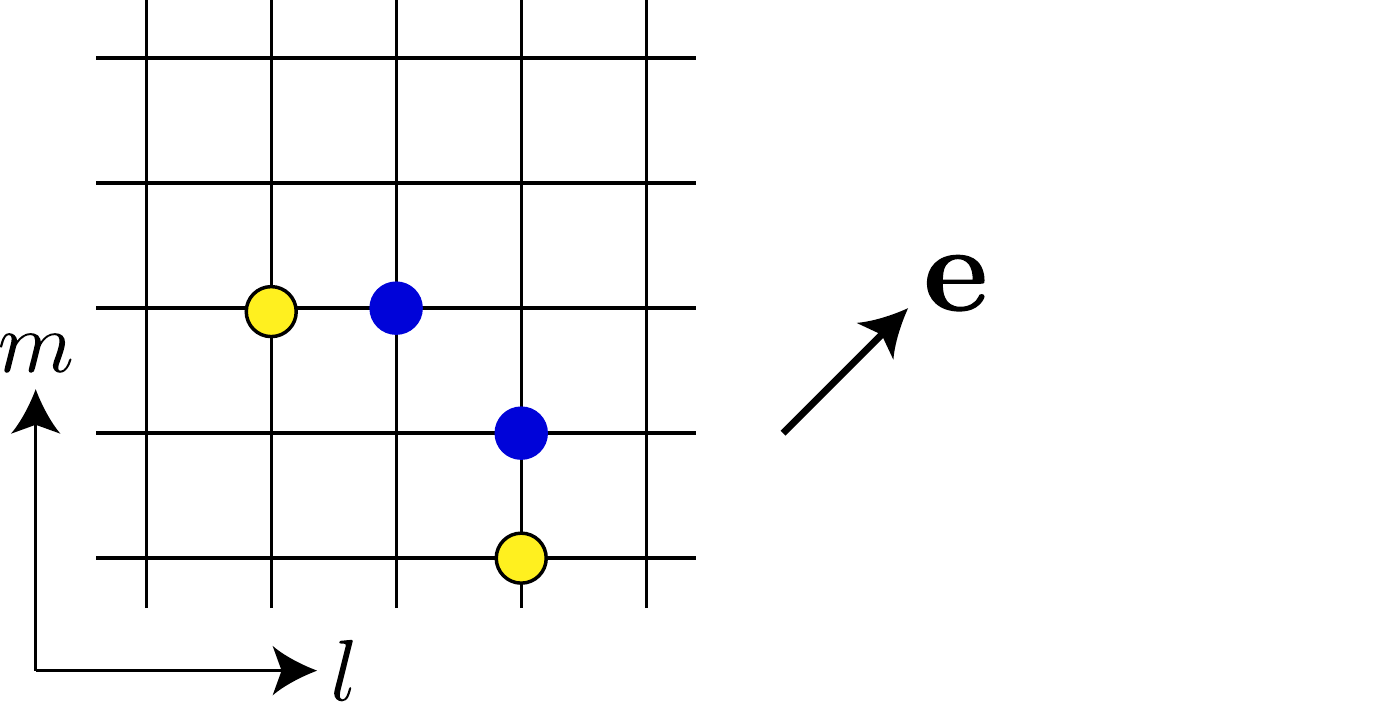}
\end{center}
\caption{(Color online) Full blue circles denote quasiparticles, and
black circles with light yellow filling denote quasiholes. The tilt vector is ${\bf e} =(1,1)$
and the resonant particle-hole excitations are as shown in the figure.
}
\label{fig1}
\end{figure}

For the diagonal tilt, we choose the vector ${\bf e} = (1,1)$ in the Hubbard model in Eq.~(\ref{hubbard}).
In this situation each site has two neighbors to which resonant tunneling is possible; see \fref{fig1}.
Here and in the following, we label the lattice sites as $(l,m)$ where $l$ ($m$) represents the $x$ ($y$) coordinate, respectively.
Using the methods of Ref.~\onlinecite{SachdevMI}, the Hamiltonian of the resonant subspace
can be written as
\begin{equation}
H_{ph} = -t \sqrt{n_0 (n_0+1)} \sum_{lm} \left[\left( \pdag_{l (m+1)}+\pdag_{(l+1) m}\right)\hdag_{lm} +\hc \right]
+\frac {\Delta}{2} \sum_{l,m}\left( \pdag_{lm}\hat p_{lm}+\hdag_{lm}\hat h_{lm}\right).
\label{eq:HSquareDiag}
\end{equation}
Here we have introduced bosonic quasiparticles $\pdag_{lm}$ and quasiholes $\hdag_{lm}$, and we identify the parent Mott insulator
with filling $n_0$, $\ket{M n_0}$, as quasiparticle and quasihole vacuum,
$\ket{0}$, and so
\begin{eqnarray}
\pdag_{l,m}\ket 0 &:=& \frac 1 {\sqrt{n_0+1}} \bdag_{l,m}\ket{M n_0},\\
\hdag_{l,m}\ket 0 &:=& \frac 1 {\sqrt{n_0}} \hat b_{l,m}\ket{M n_0}.
\end{eqnarray}
These operators are hard-core bosons and so we have the constraints
\begin{subequations}
\begin{eqnarray}
\pdag_{l,m}\hat p_{l,m}&\le& 1,\\
\hdag_{l,m}\hat h_{l,m}&\le& 1,\\
\pdag_{lm}\hat p_{lm}\hdag_{lm}\hat h_{lm}&=&0.
\end{eqnarray}
\label{constraint}
\end{subequations}
We are now interested in describing the global ground state of the effective Hamiltonian
in Eq.~(\ref{eq:HSquareDiag}) while it is subject to the constraints in Eqs.~(\ref{constraint}).
It is useful to first consider the limits of $\lambda \rightarrow \infty$ and $\lambda \rightarrow -\infty$, followed
by a general discussion.

\subsection{Limit $\lambda\rightarrow \infty$}
To zeroth order in $ 1/ \lambda$, the unique ground state is the vacuum, the parent Mott insulator.
All particle-hole excitations are gapped with the same energy (see \fref{fig1}), $E_{1} =  \Delta$.
At second order in $1/ \lambda$, we obtain an effective Hamiltonian for particles and holes, whose
structure is strongly constrained by \eqnref{constraint}. A number of distinct processes appear in the perturbation
theory:
\begin{figure}
\begin{center}
\includegraphics[width=3in]{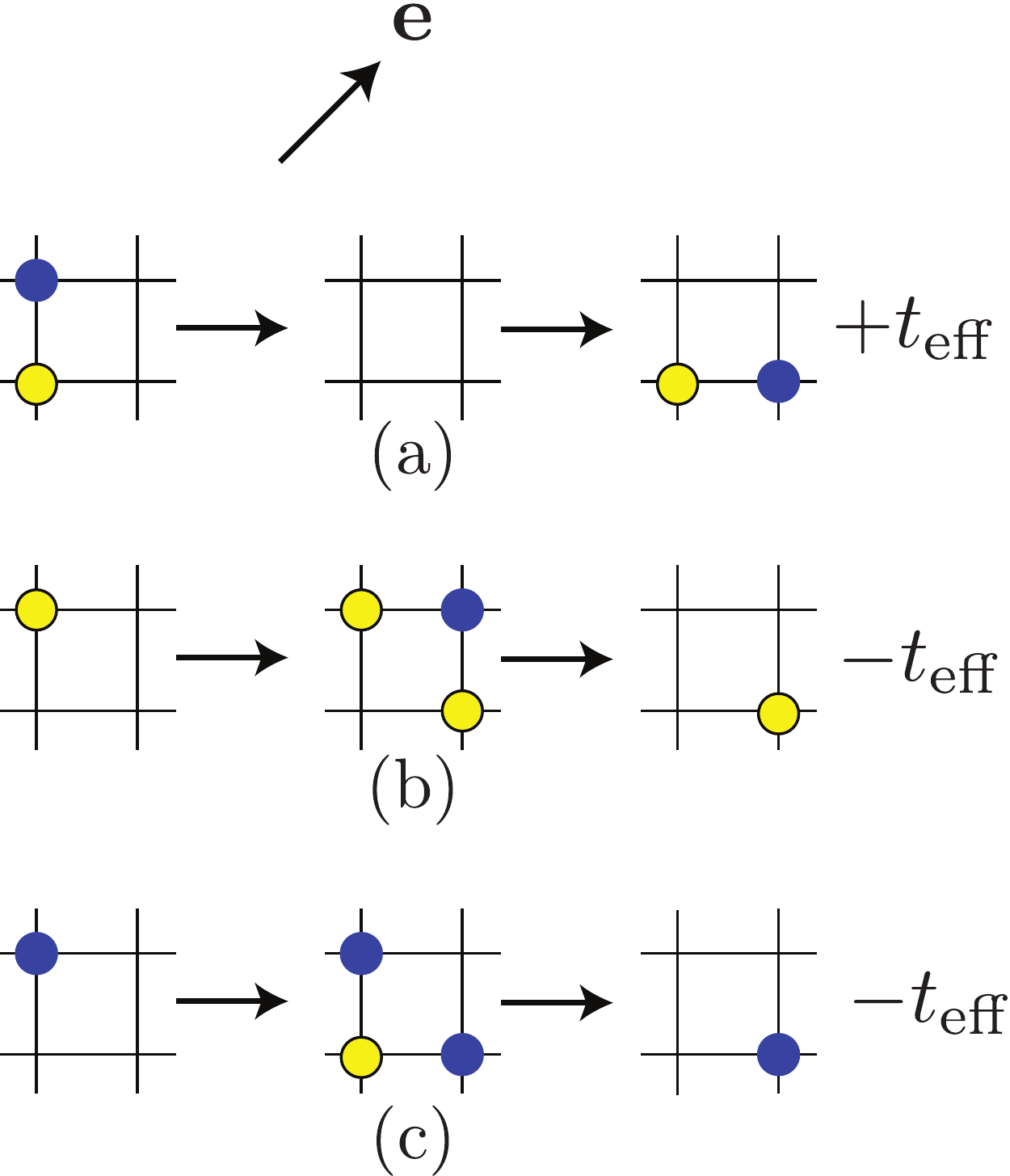}
\end{center}
\caption{(Color online) The matrix elements for the above processes are $+t_{\textrm{eff}}$ for process (a), and $-t_{\textrm{eff}}$ for (b) and (c).}
\label{fig:hopping}
\end{figure}
\begin{itemize}
\item Overall energy shift due to virtual processes. The vacuum couples to all states with one neighboring particle-hole pair, and so the vacuum energy is shifted down. Each particle-hole state couples to zero and two particle-hole states. To second order the energy $E_0$ for the vacuum and $E_1$ for states with one particle hole pair are:
\begin{eqnarray}
E_{0} &=& - 2 N\frac {t^2}{\Delta}  n_0(n_0+1),\\
E_1 &=& = \Delta - (2 N-8) \frac {t^2}{\Delta}  n_0(n_0+1).
\end{eqnarray}
where $N$ is the number of lattice sites.
\item Hopping of particles and holes along direction transverse to the tilt:
Particles and holes  can hop {\em individually\/} via second order processes along lines which are orthogonal to the tilt direction, see \fref{fig:hopping}b,c.
The magnitude of the effective hopping is
\begin{equation}
\teff=\frac {t^2 n_0(n_0+1)}{\abs{\Delta}}.
\end{equation}
It is this process which leads to transverse superfluidity. Of course, because the superfluidity is now one-dimensional, it is only quasi-long range, and characteristic of a Luttinger liquid.
\item When particles and holes are proximate to each other, we should consider their hopping
as contributing to the hopping of a dipole particle-hole pair. One such process is shown in Fig.~\ref{fig:hopping}a,
and it leads to the motion of a particle (or rotation of a dipole) with matrix element opposite to that without
an adjacent hole.
\end{itemize}
Collecting these processes, we can write down an effective Hamiltonian in the
manifold of excited states with one particle hole pair (with energy $\approx \Delta$). We label ($l_p$, $m_p$) the position of the quasiparticle, and  ($l_h$, $m_h$) the position of the hole. Clearly the vacuum state only couples to states where particle and hole are on neighboring transverse diagonals, so that $m_p+l_p=m_h+l_h+1=d$. We only need three integers to describe the positions of particle and hole: $d, m_p, m_h$. We label the states by
\begin{equation}
\ket{d,m_p,m_h}=\pdag_{d-m_p,m_p}\hdag_{d-m_h-1,m_h}\ket 0
\end{equation}
and the effective Hamiltonian in this manifold is
\begin{eqnarray}
&& H_{\textrm{ph, eff}} = E_1+
\teff \sum_{d,m}
\Big[
\left(\ket{d,m,m}+\ket{d,m,m-1}\right)
\left(\bra{d+1,m,m}+\bra{d+1,m+1,m}\right)\nonumber\\
&& \quad \quad +\ket{d,m,m}\bra{d,m+1,m}+\ket{d,m,m}\bra{d,m,m-1}+\hc\Big]\nonumber\\
&&~-\teff\sum_{d,m_p\ne m_h}\Big[
\ket{d,m_p,m_h}\bra{d,m_p+1,m_h}+
\ket{d,m_p,m_h}\bra{d,m_p,m_h-1}
+\hc\Big].
\end{eqnarray}
Here the first sum is for hopping of particle-hole pairs, both in transverse and longitudinal direction of the tilt. The second sum is for particles and holes hopping individually in transverse direction.
Diagonalization of this Hamiltonian, as in Ref.~\onlinecite{SachdevMI}, yields a continuum of separated particle-hole
excitations, along with a dipolar particle-hole bound state.

\subsection{Limit $\lambda\rightarrow-\infty$}

\begin{figure}
\begin{center}
\includegraphics[width=4in]{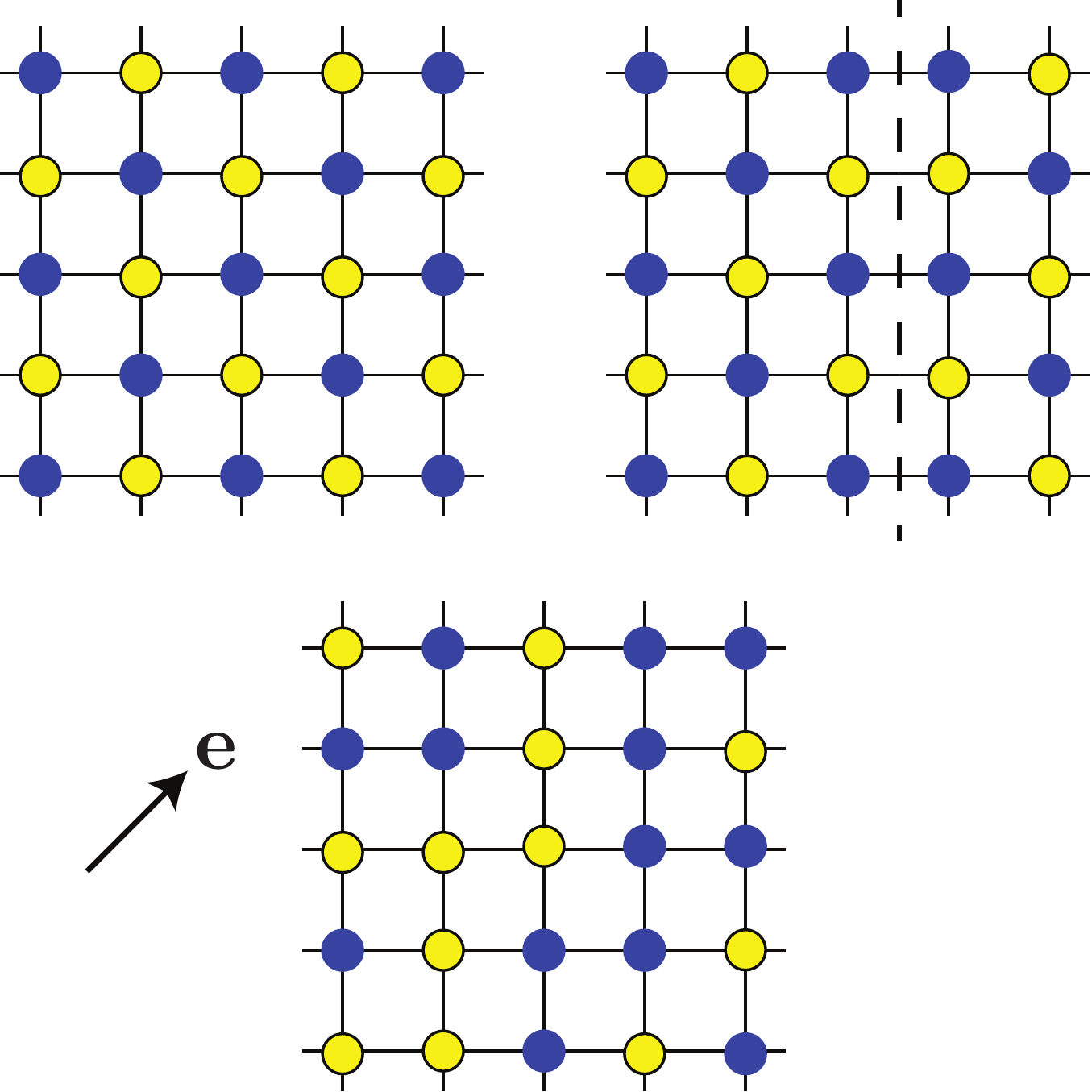}
\caption{(Color online) $\lambda\rightarrow -\infty$: Three possible states with maximum dipole density. At zeroth order in $1/|\lambda|$,
these states are degenerate, along with an infinite number of other state with the same dipole density.
At second order in $1/|\lambda|$, two states are selected as the global ground states: the checkerboard state
on the top-left, and its symmetry-related partner. The state in the top-right has a domain wall between the
two checkerboard states: the domain wall costs energy $ {t^2}/{(2\abs{U-E})}$ per unit length.
}
\label{fig:lambdainfty}
\end{center}
\end{figure}

At zeroth order in $ 1 / {\abs{\lambda}}$
there are an infinite number of degenerate ground states, all maximizing the number of particle-hole pairs.
Three such states are shown in \fref{fig:lambdainfty}.
In Ref.~\onlinecite{SachdevMI}, it was stated that these degenerate ground states can
be labelled by the set of dimer coverings of the square lattice: this is incorrect, because given a set of particle
and hole configurations, there is, in general, no unique assignment of them into nearest-neighbor
dipoles.

We can examine the lifting of the ground state degeneracy in a perturbation theory in $1/|\lambda|$.
At leading order, the matrix elements in the ground state subspace are all diagonal.
Each particle that has one hole to the left or below, can undergo a virtual annihilation, reducing the energy by
$({t^2}/{\abs{\Delta}}) n_0 (n_0+1)=\teff$.
If it has two neighbors that are holes (left and below) it can do two virtual hopping processes, reducing its energy by $2\teff$. Thus we should optimize the latter configurations: this leads to two  degenerate ground states, which look like checkerboards. These states break a lattice translation symmetry, and there is an Ising order parameter.

\subsection{Discussion}
The basic phenomenology that has emerged from our discussion so far of the diagonal tilted square
lattice is quite similar to that for the principal axis tilt considered in Ref.~\onlinecite{SachdevMI}.
For $\lambda \rightarrow \infty$, we have continua of particle-hole excitations, while
for $\lambda \rightarrow -\infty$ we have a state with Ising density wave order,
and domain wall excitations.
While for the principal axis tilt,
these results appeared already at first order in $1/|\lambda|$, here for the diagonal tilt case we had
to go to order $1/\lambda^2$ to obtain the free hopping of particles and holes, and the appearance of the Ising order.
We don't expect these distinctions to be important at smaller values of $|\lambda|$, and so the phase
diagram of the diagonal tilt case should be similar to that of the principal axis tilt: Ising order appearing
as $\lambda$ decreases, along with transverse quasi-superfluidity 
at intermediate values of $\lambda$.

\section{Triangular lattice}
\label{sec:triangular}

As in Section~\ref{sec:square}, this section also assumes $\abs{U_{3}} \gg t$.
%
The triangular lattice case is very similar to the square lattice case. The phase transitions are to Ising
density-wave ordered states and possibly transverse quasi-superfluid phases. The tilt breaks the rotation symmetry, so the dipole states on the tilted triangular lattice are not frustrated.

\subsection{Tilt along a principal lattice direction}
\label{sec:triangA}

\begin{figure}
\begin{minipage}{.4\textwidth}
\includegraphics[width=.9\textwidth]{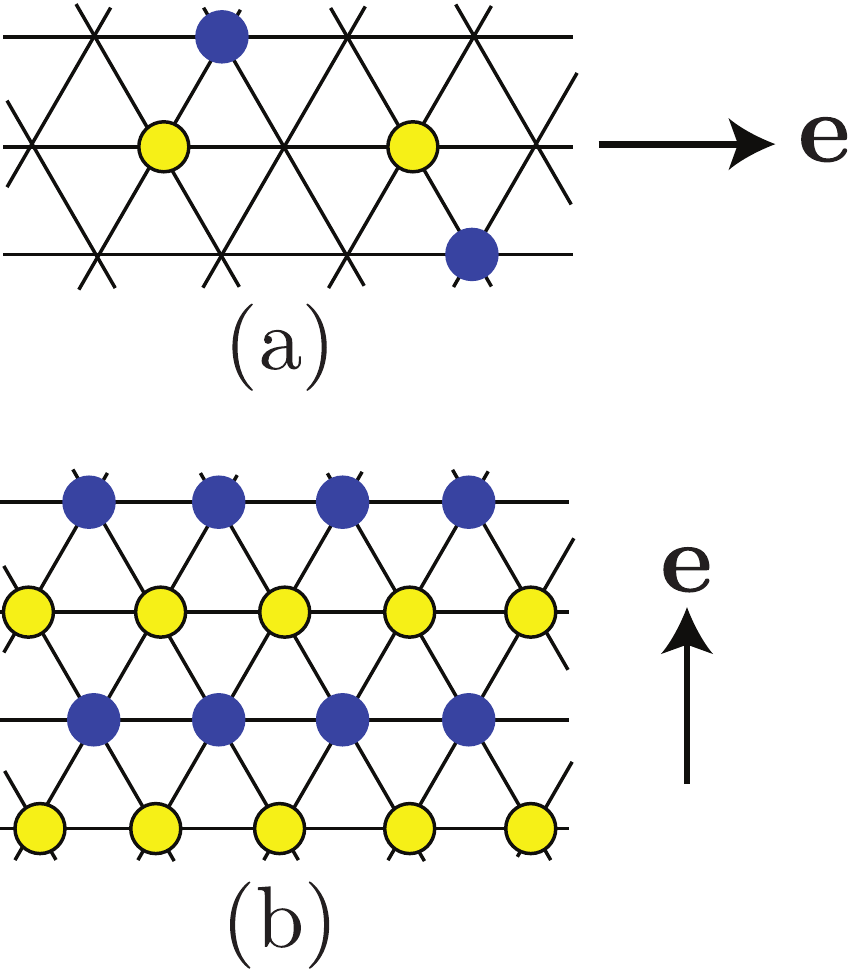}
\end{minipage}
\caption{(Color online) (a) Triangular lattice tilted along principal lattice direction. The effective Hamiltonian of the resonant subspace
for this system is the same as the one for a square lattice with diagonal tilt. Full blue circles denote quasiparticles, and black circles with yellow filling denote quasiholes. (b) Triangular lattice tilted perpendicular to a principal lattice direction.
For large negative $\lambda$, there are two-fold degenerate ground states with broken lattice symmetry.}
\label{fig:triangA}
\end{figure}

For the tilt along a lattice direction, e.g. $\vec a_1$, we choose the tilt magnitude so that the creations of dipoles along the two other lattice directions, $\vec a_2$ and $\vec a_3$, are resonant (see \fref{fig:triangA}a) but the dipole creation along the direction $\vec a_1$ is {\it not}.
Therefore, the effective Hamiltionian in the resonant subspace for this case is the same as for the square lattice with a diagonal tilt, and all conclusions from \secref{sec:square} apply here.

One could also choose the tilt magnitude so that processes along $\vec a_1$ are resonant. Then all processes along $\vec a_2$ and $\vec a_3$ are off-resonant, and the resonant subspace separates into decoupled one-dimensional systems.

\subsection{Tilt perpendicular to a principal lattice direction}
\label{sec:triangB}

Here we briefly consider a triangular lattice with tilt along a principal direction.
In the limit $\lambda\rightarrow\infty$, the ground state is the parent Mott insulator and all
excitations are gapped.
Particles and holes can separate and hop along the axis transverse to the tilt direction and reduce their kinetic energy.

On the other hand, for large and negative $\lambda$, there are precisely
two degenerate ground states, as illustrated in \fref{fig:triangA}b.
There is an Ising density wave order parameter associated with these states.
These states and their symmetries are similar to those found for the tilted square lattice
in Ref.~\onlinecite{SachdevMI}, and a similar phase diagram is expected.

\section{Decorated square and kagome lattices}
\label{decorated_lattices}

As in Sections~\ref{sec:square} and~\ref{sec:triangular}, this section also assumes $\abs{U_{3}} \gg t$.
We will now consider frustrated models in the sense that in the strongly tilted limit, $\lambda\rightarrow-\infty$, not all sites can participate in dipole creation.
This requires lattices with a larger unit cell, as will become clear from our discussion.
Furthermore, the tilt direction of those lattices can be chosen such that  independent motion of particles and holes is not possible in the resonant subspace.
We can then label the resonant subspace by a set of dipole coverings.
In the limit $\lambda \rightarrow - \infty$, we will obtain a large degeneracy in the dense dipole states.
This degeneracy is lifted by corrections in inverse powers of $|\lambda|$.
The leading corrections are not diagonal in the basis of dipole coverings:
this sets up the possibility for novel quantum liquid phases.

\subsection{Decorated square lattice with a diagonal tilt}
\label{sec:dsquare}

\begin{figure}
\begin{center}
    \includegraphics[width=2.6in]{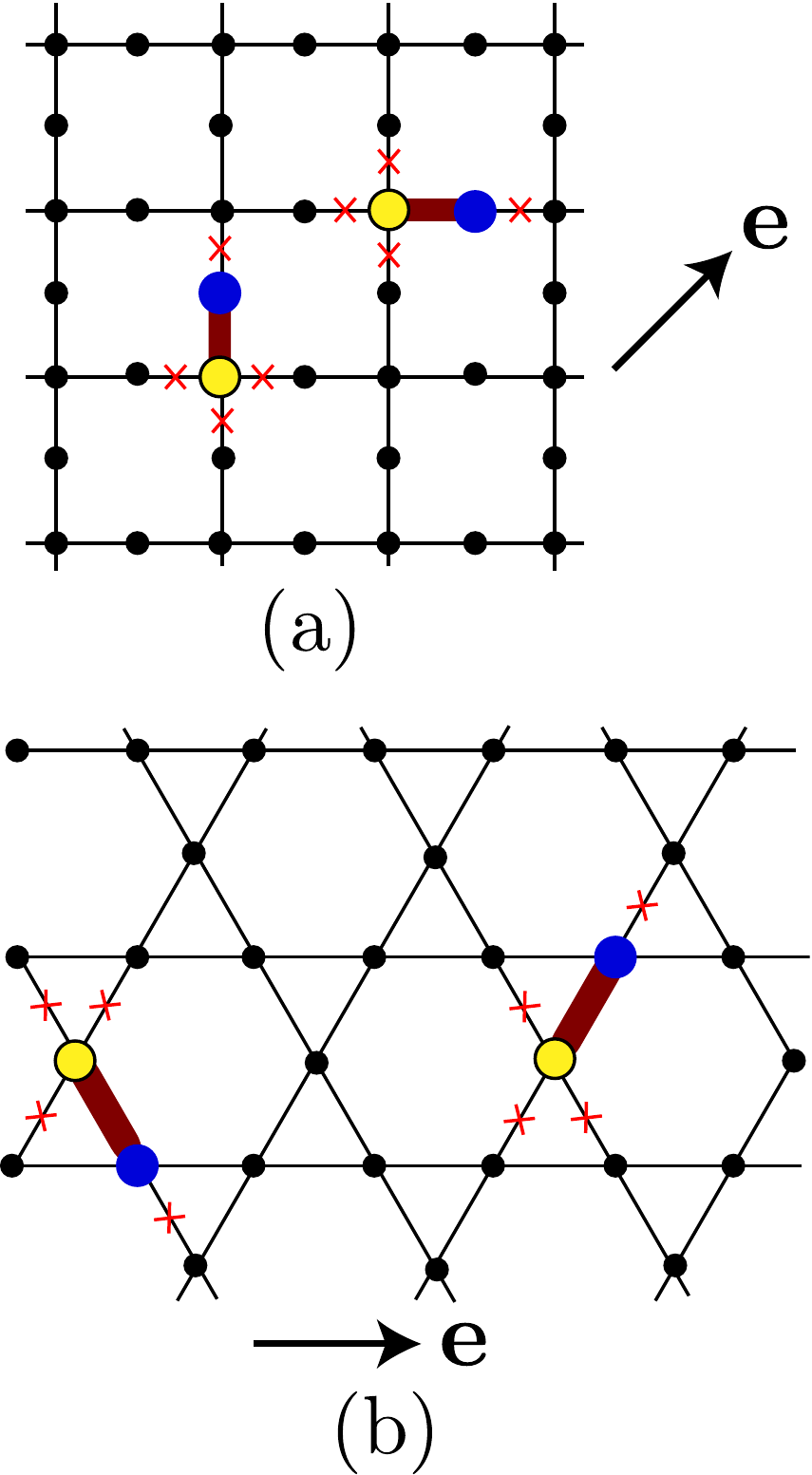}
\end{center}
\caption{(Color online) (a) Decorated square lattice in diagonal tilt, and (b) kagome lattice with tilt along lattice direction, where the tilt strength is chosen so that only processes along the diagonal lines are resonant. These two models lead
to the same effective Hamiltonian in the resonant subspace.
 Excitations of the parent Mott insulator
 are dipoles (shown as think lines connecting quasiparticle and quasihole). The particles
 and holes cannot separate. Each dipole link blocks four other links from forming dipoles (shown as
 crosses). Those are also the four links onto which each dipole is allowed to hop through virtual processes.
 }
\label{fig:diluted_square}
\end{figure}

Here we consider a decorated square lattice tilted along a diagonal direction, as illustrated in \fref{fig:diluted_square}a.
We define the distance between nearest-neighbor sites to be $a$ so that
the unit cell of the square lattice in \fref{fig:diluted_square}a has size $2a$.

Interestingly, the effective Hamiltonian for the resonant subspace resulting from the tilted
kagome lattice shown in \fref{fig:diluted_square}b is the same as the one obtained in the
decorated square lattice in  \fref{fig:diluted_square}a. While the geometry of the kagome lattice
is different from the decorated square lattice, for the specific tilt of \fref{fig:diluted_square}b,
dipoles cannot be resonantly created along the links parallel to the tilt direction.
This reduces the geometry of the tilted kagome to the tilted decorated square lattice.
Using this identity, we will present most of the results using the decorated
square lattice illustrated in \fref{fig:diluted_square}a.

The decorated square lattice has 3 sites per unit cell. Let $N$ be the number of unit cells, and we assume periodic boundary conditions. There are two kinds of sites on this lattice: $N$ sites that have four neighbors, and 2$N$ sites that only have two neighbors. We define the unit cell so that a site with four neighbors is in its center. We will refer to these sites as central sites.

The Hamiltonian in the resonant subspace is
$$
H = \Delta \sum_{a} \ddag_a \hat d_a -t \sqrt{n_0(n_0+1)} \sum_{a}\left(\hat d_a +\ddag_a\right),
$$
where $a$ labels the lattice links, and $\ddag_a$ creates a dipole on this link. The resonant subspace
has the constraints
\begin{equation}
\ddag_a\hat d_a\le 1,
\label{dimerhardcore}
\end{equation}
and
\begin{equation}
\ddag_a \hat d_a \ddag_{a^\prime} \hat d_{a^\prime} =0
\label{dimersconstraint}
\end{equation}
if $a$ and $a^\prime$ share a lattice site, see \fref{fig:diluted_square}.
Note that particles and holes cannot separate on this lattice through hopping in the resonant subspace.
It is notable that this effective model restores the rotational symmetry of the lattice and is symmetric under
rotation of $\pi/2$ around the central sites.

\subsubsection{Limit $\lambda\rightarrow\infty$}

\begin{figure}[tb]
\begin{center}
    \includegraphics[width=3in]{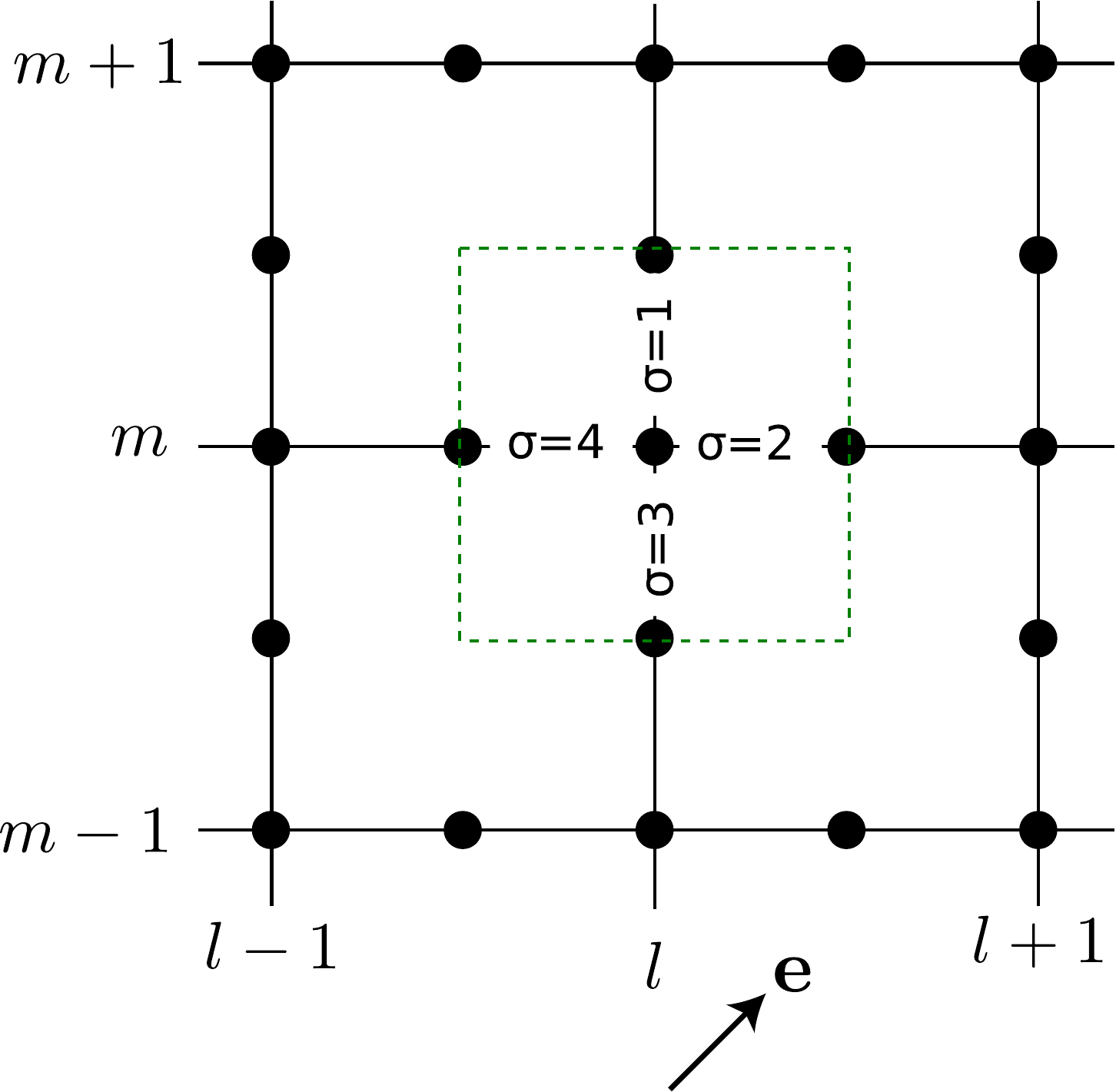}
\end{center}
\caption{(Color online) Labeling of sites and links on the decorated square lattice.}
\label{fig:unitcell}
\end{figure}

Here the ground state is the parent Mott insulating state. The excitations are gapped and
the lowest excitations are given by the creation of a single dipole.
To zeroth order in $ 1 /\lambda$, all states with one dipole are degenerate.
We now describe the effective Hamiltonian for the manifold of excited states with energy $\approx\Delta$ to second order in $1/ \lambda$. First of all, there are energy shifts for the vacuum energy $E_0$ and energy of one dipole $E_1$, given by
\begin{eqnarray}
E_0&=&-4N\teff\\
E_1&=&\Delta-(4N-6)\teff
\end{eqnarray}
with
\begin{equation}
\teff=\frac{t^2 n_0 (n_0+1)}{\abs{\Delta}}.
\label{eq:teff}
\end{equation}
Additionally, second order processes lead to
hopping of the dipoles. Note that a dipole always involves one central site and one neighboring
site located to the right, left, up or down relative to the central site.
Thus, we label the dipole states by $\ket{l,m,\sigma}$, where $(l,m)$ denotes the position of the central
site and $\sigma=1,2,3,4$ indicates the direction of the dipole, see \fref{fig:unitcell}.
With this notation, the effective Hamiltonian for one dipole excitation is given by
\begin{eqnarray}
H_{\rm 1 dimer}&=& E_1+\teff \sum_{\sigma\ne \sigma^\prime,m,l}
\ket{l,m,\sigma}\bra{l,m,\sigma^\prime}\nonumber\\
&&+\teff\sum_{l,m}
\left[
\ket{l,m,1}\bra{l,m+1,3}+\ket{l,m,2}\bra{l+1,m,4}+\hc
\right]. \label{eq:dimerham}
\end{eqnarray}
The first sum represents hopping of the dipole at a central site from one orientation to another, and the second sum represents hopping from one unit cell to another.
Diagonalizing the above Hamiltonian \eqnref{eq:dimerham}, one obtains the spectra of a single dipole excitation
given by
\begin{eqnarray}
\epsilon_1 (k_x,k_y)&=&E_1-2t_{\rm eff}\\
\epsilon_2 (k_x,k_y)&=&E_1\\
\epsilon_3(k_x,k_y)&=&E_1+\left(1-\sqrt{5+2 \cos(2a k_x)+2 \cos(2a k_y)}\right)t_{\rm eff}\\
\epsilon_4(k_x,k_y)&=&E_1+\left(1+\sqrt{5+2 \cos(2a k_x)+2 \cos(2a k_y)}\right)t_{\rm eff}
\end{eqnarray}
where $2a$ is the distance between two neighboring central sites.
The first two bands are `flat bands' as they do not depend on momentum.
\subsubsection{Limit $\lambda\rightarrow-\infty$, quantum liquid state}
\label{sec:clock}
\begin{figure}[tb]
\begin{center}
\includegraphics[width=3.2in]{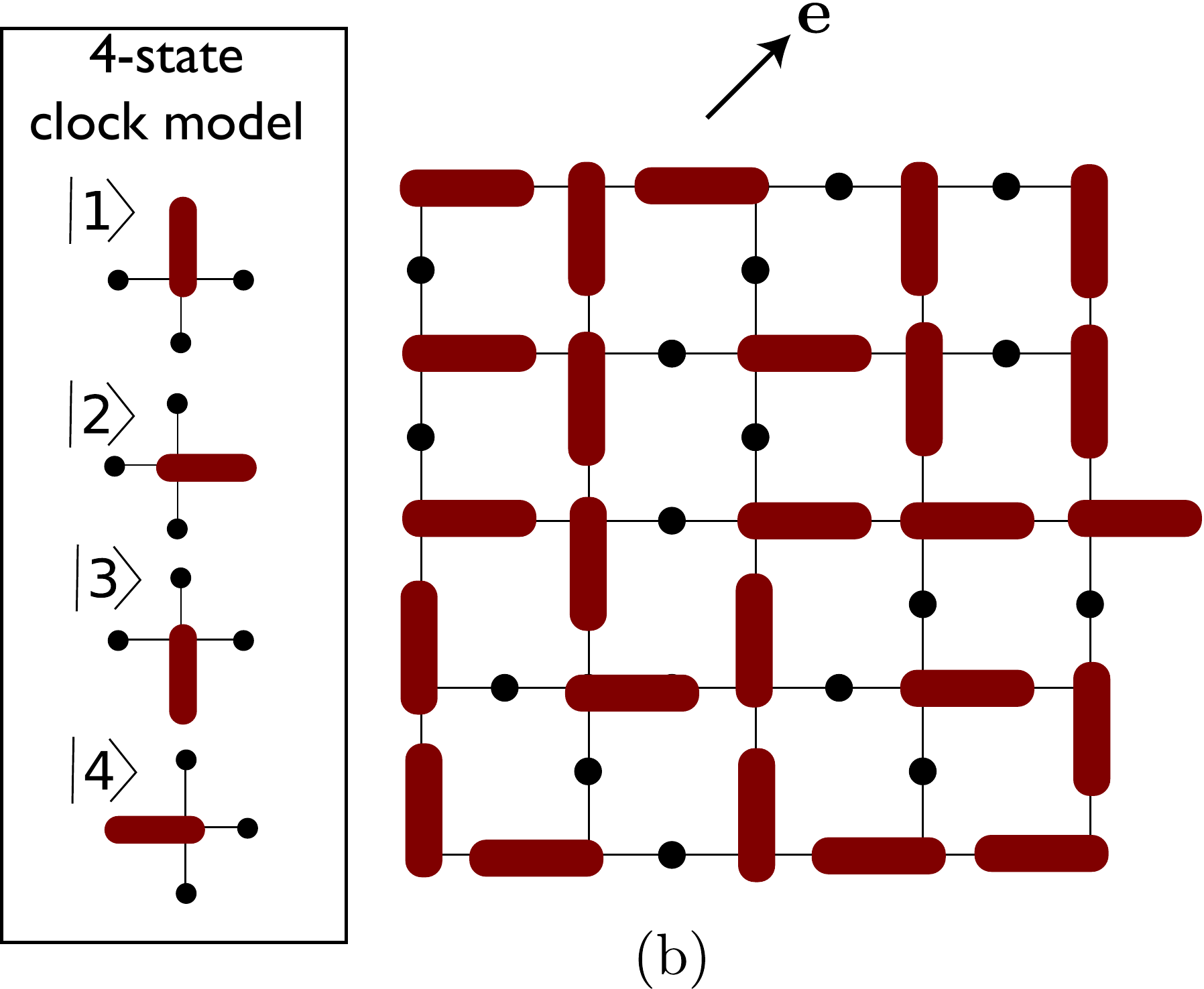}\\~\\
\includegraphics[width=1.8in]{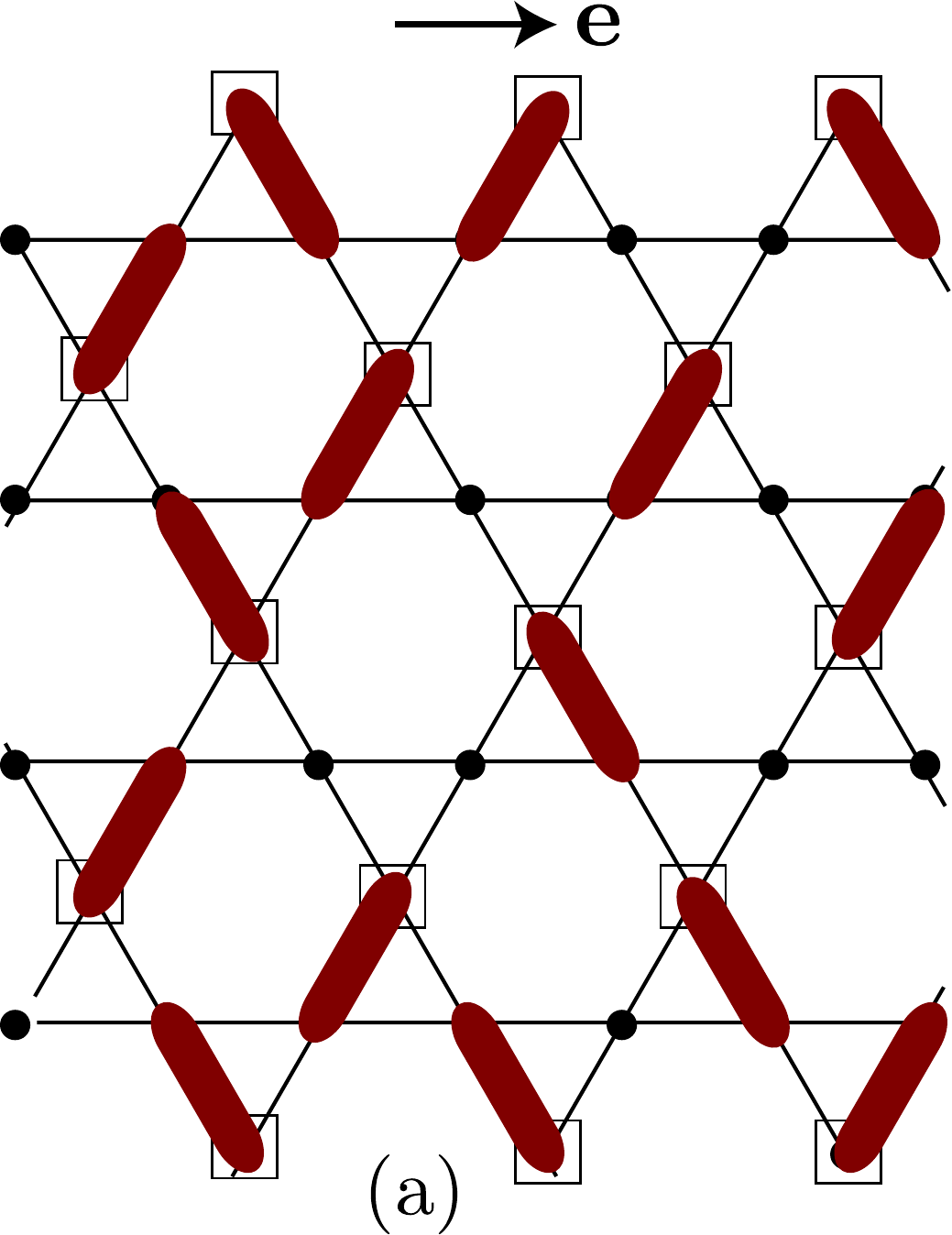}
\end{center}
\caption{(Color online) $\lambda\rightarrow-\infty$: quantum 4-state clock model: (a) decorated square lattice, (b) kagome lattice.
Both lattices lead to the same effective Hamiltonian. Minimum of the energy for the effective Hamiltonian is
obtained when the number of dipoles is maximum. This requires, on the square lattice,
every central site to have a dipole, and on the kagome lattice, every site marked with a square to have a dipole.
As a result, we can look at the system in the limit $\lambda\rightarrow-\infty$
as a collection of four state clocks, one each on the central sites of the square lattice,
or on the marked sites of the kagome.
The states on neighboring clocks are constrained by the requirement that the dipoles cannot overlap.}
\label{dimer}
\end{figure}
To zeroth order in $1 /\lambda$,
all states which maximize the number of dipoles are degenerate, and they are all ground states of the effective Hamiltonian.
These states require every central site on the square lattice to have a dipole (on the kagome lattice, every site marked with a square has a dipole), as illustrated in \fref{dimer}.
There is an exponentially large number of such states which satisfy this requirement.

To second order in $1 /|\lambda|$, dipoles can hop as long as the constraint \eqnref{dimersconstraint}
is satisfied.
As there is a dipole on every central site, the constraint implies that dipoles may only change their orientation while remaining at the same central site.
Thus, we can write the hopping of the dipoles as a 4-state quantum clock model,
with the clocks residing on the vertices of the square lattice.
The hopping Hamiltonian can be written as (see \fref{dimer})
\begin{eqnarray}
H_{\textrm{clock}} &=& \sum_{lm} H_{l,m} \nonumber \\
H_{l,m} &=& \Delta -\teff (\ket{1} + \ket{ 2}+ \ket{ 3}+\ket{ 4}) (\bra{1} + \bra{ 2}+ \bra{ 3}+\bra{ 4})
\label{eq:liquidhamiltonian}
\end{eqnarray}
where $(l,m)$ is the site index, and $\ket{\sigma}$ is a shorthand for $\ket{l,m,\sigma}$ representing
the orientation of the dipole sitting at $(l,m)$.
It should be remembered that states within the resonant subspace
have the important constraint that two dipoles sitting at neighboring central sites
cannot be directed toward each other. Such constraints are not contained in $H_{\textrm{clock}}$.
If we define the projection operator onto the resonant subspace which satisfies the constraints as $P$,
the effective Hamiltonian with the constraints is given by projecting the hopping Hamiltonian,
\begin{equation}
H^{c}_{\textrm{clock}} = P H_{\textrm{clock}} P.
\end{equation}

In the following we will show that the {\em exact} ground state of the Hamiltonian with constraint $H^{c}_{\textrm{clock}}$ is obtained through
 the projection of the ground state of the Hamiltonian without constraint, $H_{\textrm{clock}}$, onto the resonant subspace which satisfy the constraints Eq.(4.1).
The ground state of the unconstraint single-site Hamiltonian~$H_{l,m}$ is given by $\ket{1} + \ket{2}+ \ket{3}+\ket{4}$ with eigenenergy~$\Delta-4\teff$, and there are three degenerate excited states%
\footnote{Interestingly, it is possible to fill the lattice with these excited states while respecting the constraint: some of the highest energy eigenstates of $H_{\textrm{clock}}$ are also eigenstates of $H^{c}_{\textrm{clock}}$.
Note that states of the form $\ket{\sigma}-\ket{\sigma^\prime}$, with $\sigma\ne \sigma^\prime$, are excited eigenstates of the single site Hamiltonian \eqref{eq:liquidhamiltonian}. These states only occupy two links. We could, for example, fill the lattice with the state $\ket{1}-\ket{2}$ on every site.
}
with eigenenergy~$\Delta$.
Therefore, the ground state of $H_{\textrm{clock}}$ is given by
\begin{eqnarray}
\ket{\psi_{0}} = \mathcal{N} \prod_{l,m} (\ket{l,m,1} + \ket{l,m,2}+ \ket{l,m,3}+\ket{l,m,4})
\end{eqnarray}
where $ \mathcal{N} $ is some normalization constant.
As we will show in the following, the exact ground state of $H^{c}_{\textrm{clock}}$ is given by
\beq \ket{\psi_{c}} = P\ket{\psi_{0}}. \label{eq:exactgs} \eeq
Any configuration with maximum number of dipoles corresponds to
a product state given by $\prod_{l,m} \ket{l, m, \sigma_{l,m}}$ where $\sigma_{l,m} =1,2,3,4$.
Then the state $\ket{\psi_{c}}$ is nothing but the equal superposition of
all the product states in the resonance subspace in the limit $\lambda\rightarrow-\infty$
i.e. all the configurations with maximum number of dipoles.

The proof proceeds in two steps. First, we will show that this nodeless state $\ket{\psi_{c}}$ is an eigenstate
of the Hamiltonian $H^{c}_{\textrm{clock}}$. Second, we will
show that
it is necessarily the
unique ground state of the Hamiltonian $H^{c}_{\textrm{clock}}$. 

In order to see that $\ket{\psi_{c}}$ is an eigenstate of $H^{c}_{\textrm{clock}}$, we check
that $H^{c}_{\textrm{clock}} \ket{\psi_{c}}$ is again the equal superposition of
all the configurations with maximum number of dipoles.
Consider a particular configuration with
maximum number of dipoles represented by the product state $\ket{M}$. We count how many different
configurations in $\ket{\psi_{c}}$ produce $\ket{M}$ after the action of $H^{c}_{\textrm{clock}}$.
This is equivalent to counting how many different state can be created from $\ket{M}$
through the action of $H^{c}_{\textrm{clock}}$.
For any $\ket{M}$, there are in total $2N$ different configurations that are connected
with $\ket{M}$ through an action of $H^{c}_{\textrm{clock}}$, where $N$ is the total number of dipoles in the system.
The decorated square lattice has in total $2N$ non-central sites, and
in a configuration with maximum number of dipoles $\ket{M}$, $N$ of them are part of a dipole and $N$ of them
are {\it not}. Now the action of $H^{c}_{\textrm{clock}}$ is a change of the orientation of a dipole at a central site,
and the action can change the configuration by hopping a dipole to these non-central sites that are not
occupied by a dipole in $\ket{M}$. Since each unoccupied non-central site is connected with $2$ central
sites, there are in total $2N$ different configurations connected with $\ket{M}$.
This consideration is true for any $\ket{M}$, so we conclude that
$H^{c}_{\textrm{clock}} \ket{\psi_{c}} = N \left(\Delta -3 \teff \right) \ket{\psi_{c}}$. Notice that
the Hamiltonian \eqnref{eq:liquidhamiltonian} contains terms which bring a configuration back
to itself, so the eigenenergy is $N \left(\Delta -3 \teff \right)$. 
Indeed $\ket{\psi_{c}} $ is an eigenstate.

Now we show that this state is necessarily the unique ground state.
Note that if we take as the basis the product states which correspond to
configurations with maximum number of dipoles, all the off-diagonal matrix element of $H^{c}_{\textrm{clock}}$ are negative.
Moreover, repeated applications of $H^{c}_{\textrm{clock}}$ can connect
any configuration with maximum number of dipoles
to any other configurations with maximum number of dipoles.
This statement is even true when the system is placed on a manifold with non-trivial
topology, such as a torus.
From these two conditions, it follows through Perron-Frobenius theorem\cite{Perron, Frobenius, assa} that the ground state
of the effective Hamiltonian \eqnref{eq:liquidhamiltonian} is the superposition of product states with strictly positive amplitudes.
Moreover, since two of such states cannot be orthogonal to each other, this ground state
is unique. Now since we found the state $\ket{\psi_{c}}$ which is a superposition of
product states with strictly positive amplitudes, it follows that $\ket{\psi_{c}}$ must be
the unique ground state of the Hamiltonian.

\begin{figure}
    \includegraphics[width=.5\textwidth]{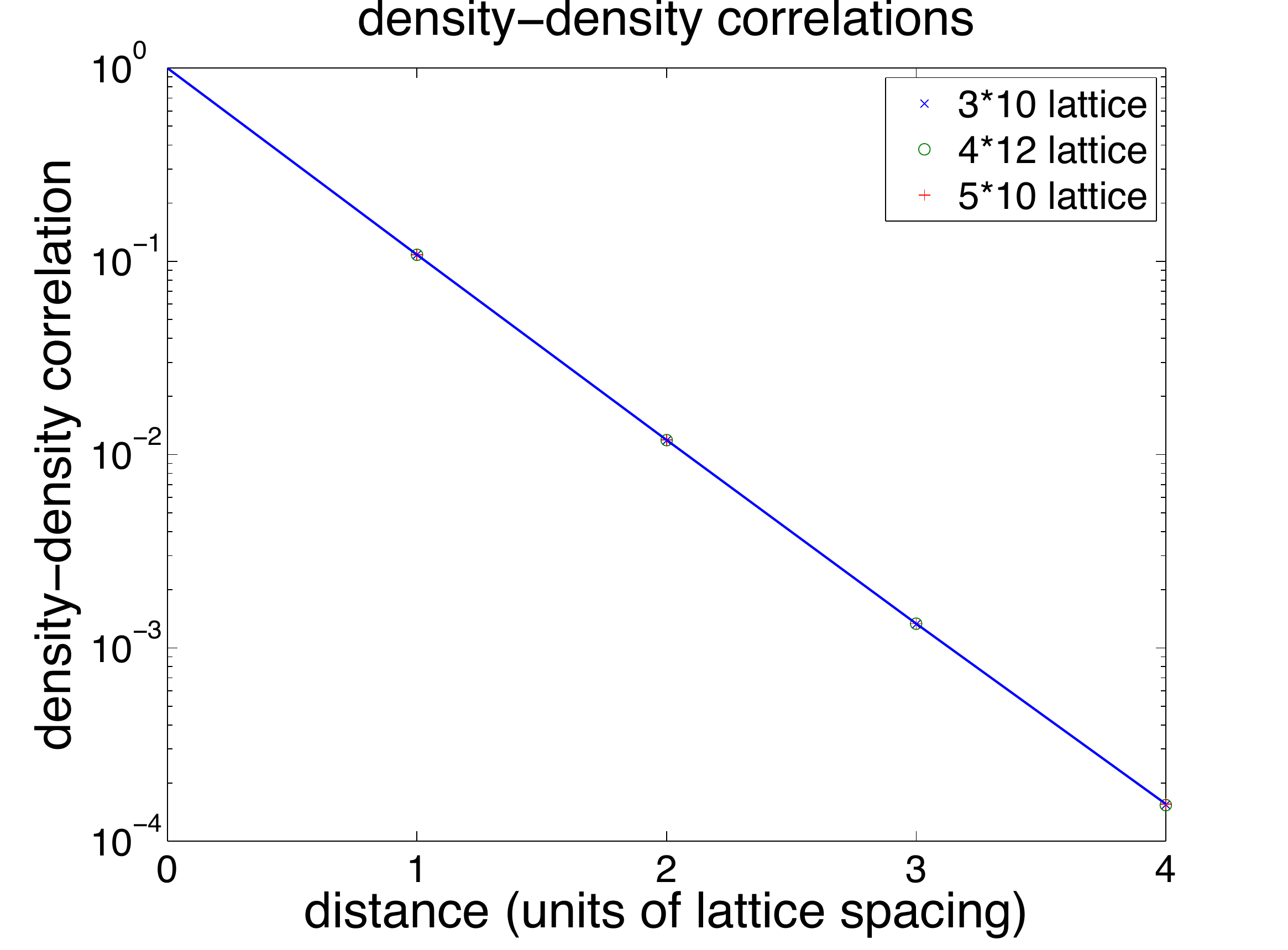}
\caption{(Color online) Density-density correlation function $\ave{\hat n_{0,0}\hat n_{0,m}}$ on the central sites (the ones marked by a square in \fref{dimer}), obtained using transfer matrices and periodic boundary conditions. Correlations decay rapidly in an exponential fashion.
A fit to the exponential $f(m)=a \exp(-bm)+c$ gives $a=0.9999$, $b=2.225$, $c=7.371\times10^{-5}$, this corresponds to a correlation length of less than half a lattice spacing.
}
\label{correlations}
\end{figure}
In order to probe the properties of the ground state $\ket{\psi_{c}}$,
we numerically computed correlations for the state in Eq.~(\ref{eq:exactgs}).
For the correlations of local operators in a state which is the equal superposition of product states,
the correlations can be related to the corresponding classical problem \cite{Kivelson88}.
Here we study the density-density correlation $\ave{\hat n_{0,0}\hat n_{0,m}}$ on the central sites. $n_{l,m}$ is the density of the central site at $(l,m)$ measured relative to the density of the Mott insulator, so that $n_{l,m}$ can, in principle, take the values $+1, 0, -1$. Here $+1$ ($-1$) means that there is an extra (a missing) boson on that lattice site, and that this site has formed a dipole with a neighbor below or to the left (above or to the right). $n_{l,m}=0$ would mean that this central site is not participating in a dipole bond; in the limit $\lambda\rightarrow-\infty$ all central sites have $n_{l,m}=\pm 1$.
We used row-to-row transfer matrices with rows of length up to~$5$ unit cells and periodic boundary conditions to compute
density-density correlations\cite{baxterexact}.
Results for lattice sizes $3\times 10$, $4\times 12$, and $5\times 10$ are plotted in \fref{correlations}. We obtain essentially the same results for different lattice sizes, and a fit of the correlation function to the exponential $f(m)=a \exp(-bm)+c$ gives $a=0.9999$, $b=2.225$, $c=7.371\times10^{-5}$, corresponding to a correlation length of less than half a lattice spacing. Similarly, dipole-dipole
correlations can also be calculated: let $\hat d_{\sigma,l,m}$ be an operator that projects onto states which have dipole orientation $\sigma$ at site $(l,m)$. We find that correlations $\ave{\hat d_{\sigma,0,0}\hat d_{\sigma^\prime,l,m}}-\ave{\hat d_{\sigma,0,0}} \ave{\hat d_{\sigma^\prime,l,m}}$ decay exponentially.

The exponential decay of the equal-time correlations suggests that the $\lambda \rightarrow - \infty$ liquid state
has a gap to all excitations. Furthermore this liquid state is non-degenerate on a torus, implying the absence of topological order, and we expect the ground state to remain unique
in the thermodynamic limit $N\rightarrow\infty$.
Thus there is a possibility that the parent Mott insulator state at $\lambda \rightarrow \infty$ is adiabatically connected
to the liquid state at $\lambda \rightarrow -\infty$, without an intervening quantum phase transition.

Ultracold atom experiments are expected to realize
this quantum liquid state
as long as the temperature is lower than the energy scale of $\teff \propto {t^2}/{|E-U|}$.
Since the magnitude of $\teff$ can be controlled by changing the tilting strength $E$, it is likely that
the temperature of the order of $\teff$ is achievable in experiments. Moreover, recent experiments
by Bakr {\it et. al.}\cite{BakrGreiner} showed that the Mott phase contains particularly low entropy
compared to the surrounding superfluid, and fits with theoretical curve suggest that the Mott phase
is as low temperature as a few nK. Therefore, together with the demonstration of lattice tilting\cite{bloch, 1dIsingExp} and
possibility to create various lattice structures through a holographic mask technique\cite{bakrgreiner2},
realization of this quantum liquid state should be
possible with current technology.

\subsection{Doubly-decorated square lattice}
\label{sec:ddsquare}

Although the quantum liquid state found in Section~\ref{sec:dsquare} has non-trivial entanglement,
it does not have topological order.
Here we show that by decorating the square lattice further, it is possible to obtain a model whose
resonant subspace is labeled by the set of dimer packings of the undecorated square lattice.
The dipole hopping terms appear in the perturbation theory of $1/\lambda$, and the effective
Hamiltonian in this resonant subspace is expected to take
a version of the quantum dimer model\cite{Kivelson88}.

\begin{figure}[tb]
\begin{center}
\includegraphics[width=1.8in]{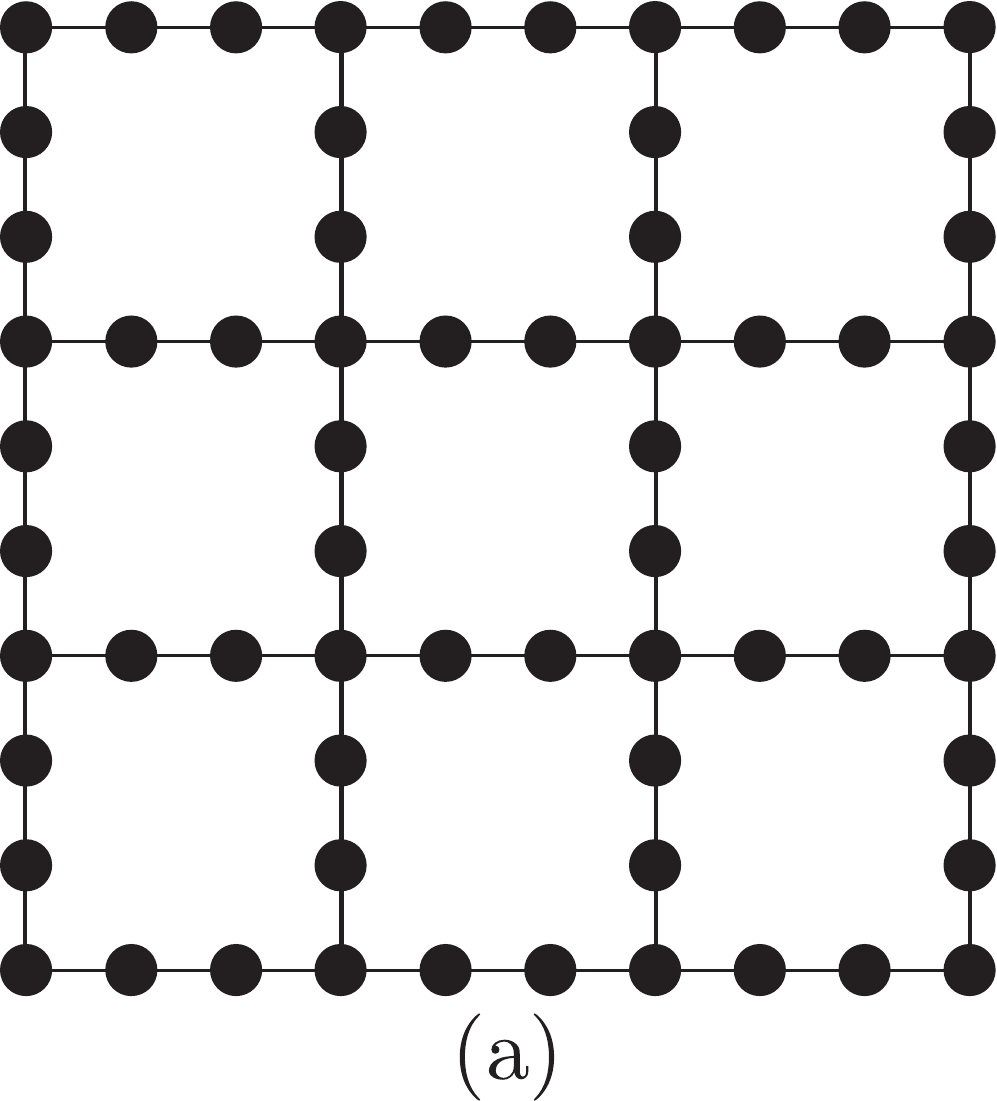}
\\  \includegraphics[width=1.8in]{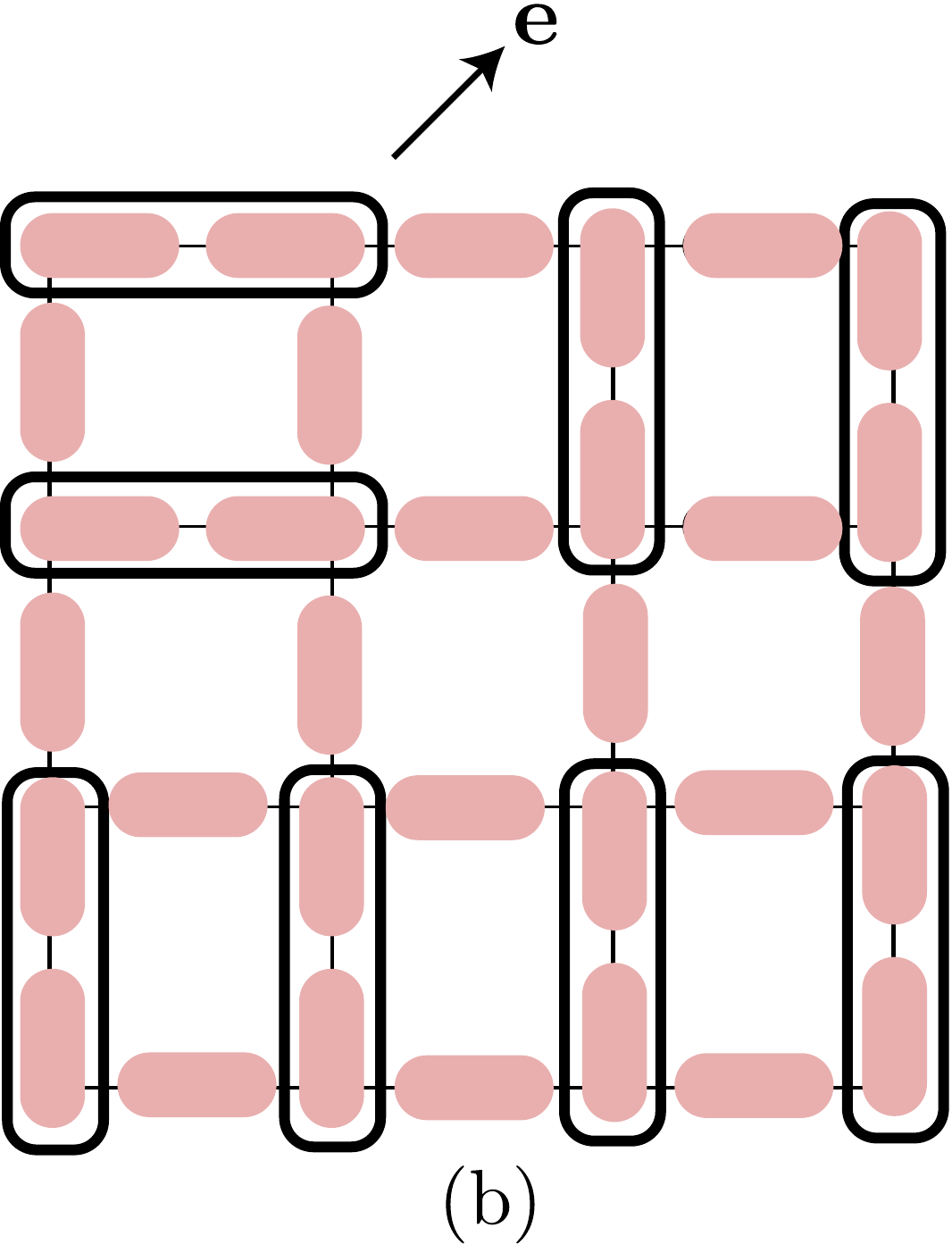}
\end{center}
\caption{(Color online) (a) The doubly-decorated square lattice. (b) Dense dipole configuration obtain in the limit $\lambda \rightarrow - \infty$.
Each dipole is represented by the thick line, and consists of a particle-hole pair. Links with two dipoles are identified
as having a ``dimer'', and these are indicated by the ovals. These dimers are described by an effective quantum dimer model.
}
\label{fig:dimermodel}
\end{figure}

In this model, we consider a square lattice decorated by {\em two\/} sites at each link between lattice sites,
as shown in Fig.~\ref{fig:dimermodel}a.
Quantum spin models on a similar decorated square lattice were considered in Ref.~\onlinecite{raman}.
In the limit $\lambda \rightarrow -\infty$, the ground states in the resonant subspace are
those with maximum number of dipoles. As before, there are a large number of such states
and they are all degenerate in this limit.
The dipoles now come in two flavors: those which have one end on a central site of the square lattice, and those which reside exclusively on a link between central sites.
Along a link between two central sites, the first kind of dipoles always come in pairs. We associate
this pair with a ``dimer'' which connects two neighboring central sites (see Fig.~\ref{fig:dimermodel}b).
Because there can be only one dimer on each site for $\lambda \rightarrow -\infty$, the dimers are close-packed, and there is a one-to-one correspondence between the degenerate ground state manifold and the set of dimer packings
of the square lattice.
Including corrections in powers of $1/|\lambda|$ 
we expect an effective Hamiltonian which acts on a Hilbert space of states labeled by the dimer coverings: this must have the form of the quantum dimer model \cite{Kivelson88}.
In particular, a
plaquette-flip term
appears at order $1/\abs{\lambda}^{12}$; up to this order, all diagonal terms in the basis of dipole coverings are equal and independent of the configurations.
We leave the details of the calculations as well as possible extensions to a future work, and give a general discussion below.

On the square lattice, quantum dimer models are generically expected
to have ground states with valence bond solid orders\cite{rsb}.
At critical points between different solid orders, topologically ordered spin-liquids can be obtained\cite{bsv, fhmo}.
{The quantum dimer model realized in this system consists of only a kinetic term, and no energy cost or gain for parallel dimers, }
\begin{equation}
H_{\rm dimer}=-t_{\rm dimer} \sum_{\rm plaquettes} \left(
\ket{||}\bra{=}+\ket{=}\bra{||}\right).
\end{equation}
This model is not critical, and so valence bond order is expected, but the precise
nature of the square lattice symmetry breaking remains under study.
An early exact diagonalization study\cite{Leung} suggested a plaquette phase, more recent quantum Monte Carlo calculations\cite{Syljuasen} predicted a columnar phase, and recently a mixed columnar-plaquette phase has been proposed\cite{Ralko}.
As $\lambda$ increases away from $\lambda = - \infty$, configurations
with less number of dipoles become energetically favorable.
These states are analogous to those in the doped quantum dimer model \cite{Kivelson88,fradkiv}, where other
novel phases are possible\cite{balents2}.
As the parent Mott insulator at $\lambda = \infty$ is non-degenerate, it cannot be connected adiabatically to all the possible $\lambda \rightarrow -\infty$ states with topological order or broken
symmetry: there must be at least one intervening quantum phase transition.

\subsection{Kagome lattice tilted perpendicular to a principal lattice axis}

\begin{figure}[tb]
\begin{center}
\includegraphics[width=2.3in]{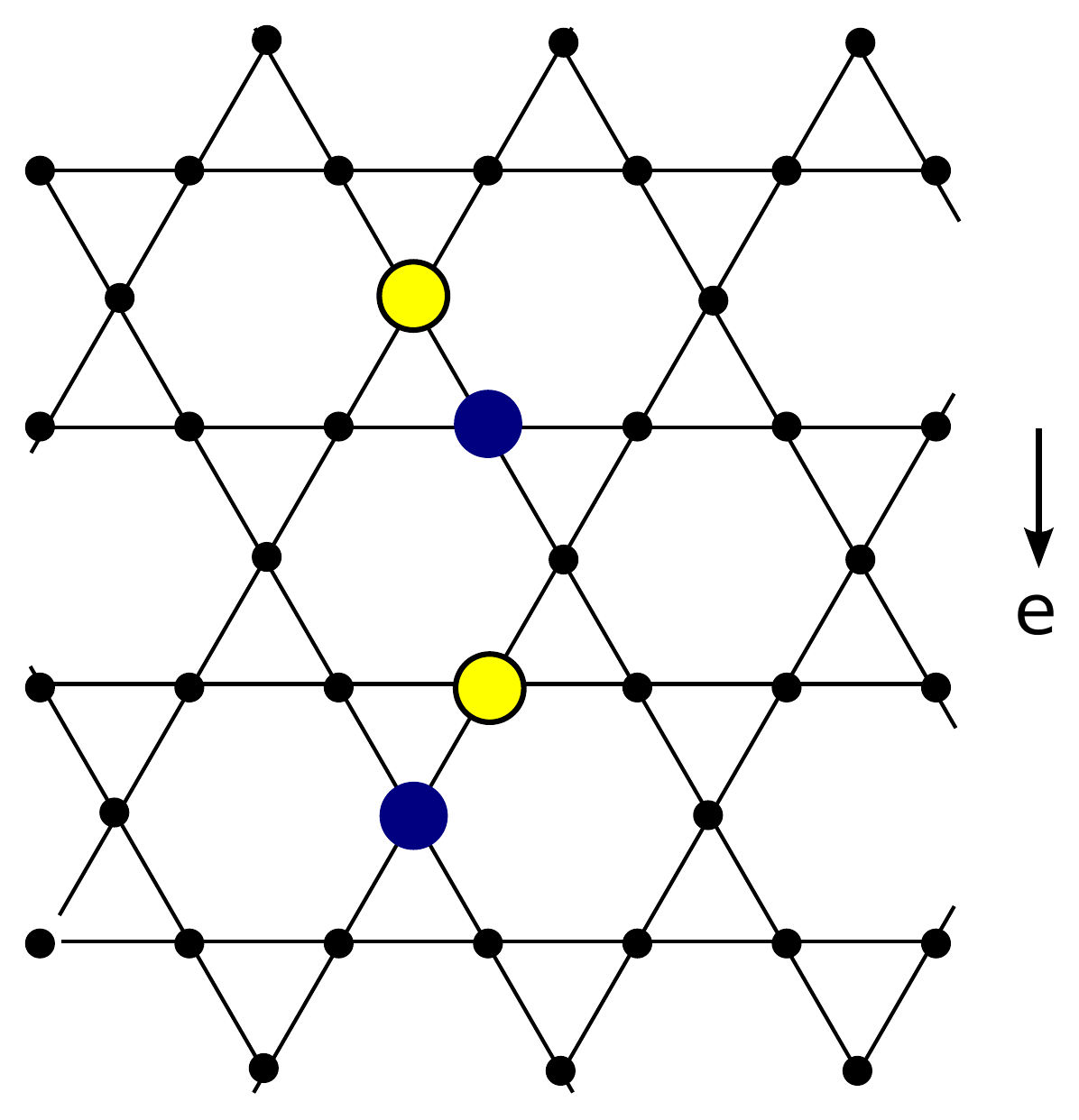}
\caption{(Color online) Kagome lattice tilted perpendicular to a principal lattice axis: resonant particle-hole excitations are illustrated. After a particle-hole pair is created,
either the particle or the hole can hop along the direction transverse to the tilt. }
\label{fig:kagome}
\end{center}
\end{figure}

At last, we study the kagome lattice tilted perpendicular to a principle axis, as illustrated in \fref{fig:kagome}.
Because of the low connectivity in the lattice structure, dipole creations are frustrated in the limit
$\lambda \rightarrow - \infty$ and even in this limit, some sites cannot not participate in the creation of dipoles.
As a result, there is no Ising ordering and the transverse superfluidity persists in the limit $\lambda \rightarrow - \infty$, as we describe below.

\subsubsection{Limit $\lambda\rightarrow \infty$}
The ground state is again the Mott insulator, and the excitations are particles and holes.
For every particle-hole pair, either the particle or the hole is located on a transverse layer where it can hop,
while the other is fixed and cannot move, see \fref{fig:kagome}.
The dispersion of the hopping quasiholes is
    $$
    \epsilon_h(k) = \Delta - 2 n_0 t \cos(ka),
    $$
while for hopping quasiparticles it is
    $$
    \epsilon_p(k) = \Delta - 2( n_0+1) t \cos(ka).
    $$
Therefore, the hopping quasiparticles have lower energy than the hopping quasiholes.
 The excitations are gapped for $\lambda\rightarrow\infty$, but excitations become gapless for some critical value~$\lambda_c$.

\subsubsection{Limit $\lambda\rightarrow -\infty$}
\label{sec:decoupled}
\begin{figure}[tb]
    \begin{center}
        \includegraphics[width=2.3in]{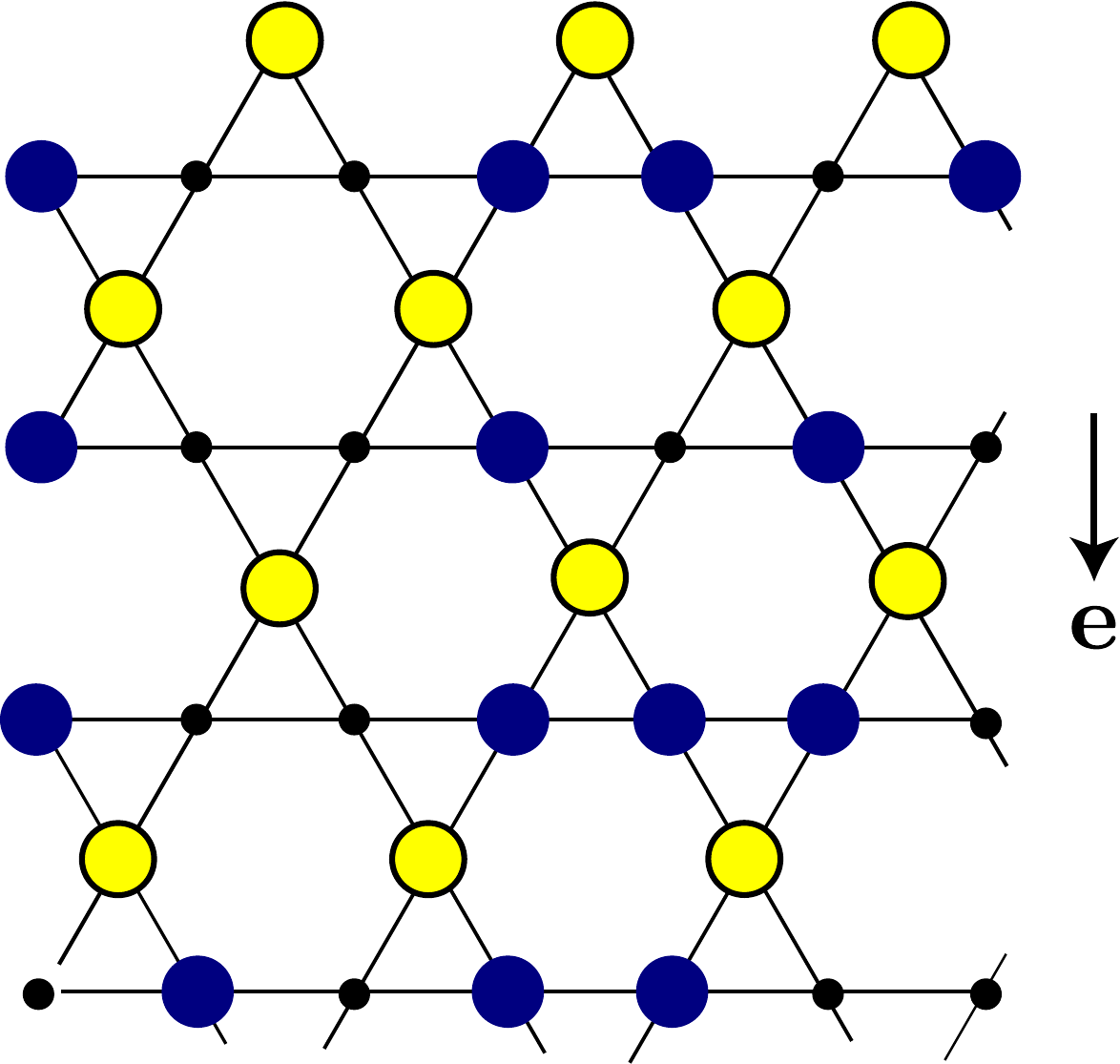}
    \end{center}
    \caption{(Color online) $\lambda\rightarrow-\infty$. Single non-degenerate ground state with half-filled Tonks gases along horizontal lines for $\lambda=-\infty$.
    }
    \label{fig:ex}
    \end{figure}

The ground state in the resonant subspace has the maximum number of dipoles.
Due to the geometry of the kagome lattice, some sites are frustrated in the sense that
they cannot participate in the creations of dipoles. These sites allow the particles and holes
to move along the horizontal direction perpendicular to the tilt,
forming Luttinger liquids. As the kinetic energy of quasiparticles is lower
than that of the quasiholes, the ground state maximizes the number of moving quasiparticles
in a transverse layer and localizes all the quasiholes, as shown in \fref{fig:ex}.
Each transverse layer thus realizes a ``Tonks-Girardeau gas'' of
bosons at half-filling\cite{girardeau}, and the ground state
realizes a collection of uncoupled gapless Luttinger liquids.
{These are gapless only in the limit $\lambda=-\infty$.
Perturbation theory to second order in $1 / \lambda$ generates a
dimerized hopping term, \emph{i.e.} the effective hopping between
two sites in a one-dimensional chain which share a neighbor above
them becomes different from
the hopping between sites which share a neighbor below. 
This breaks the translation symmetry along the one-dimensional
layers explicitly, doubles the unit cell, and opens a gap, as the
Tonks gases are exactly at half filling. Therefore, to order
$1/\lambda^2$ 
the one-dimensional systems remain decoupled, but they are no
longer gapless. }



\section{Modifications for $\abs{U_3} \ll t$}
\label{sec:tetris}
In this section we briefly discuss how the physics of tilted lattices is modified if the $U_3$ term in \eqnref{eq:repulsion} vanishes.
In this case all processes $(n,n)\rightarrow(n-1,n+1)$ for general $n$ are resonant at the same tilt magnitude. This greatly increases the size of space of states resonantly connected to the parent Mott insulator. As we will show in some cases the energy of this enlarged space is not bounded from below, and we then cannot use equilibrium methods to describe the physics. In other cases 
the resonant subspace
remains bounded from below, but the phases differ from the ones found for $\abs{U_3}\gg t$.
We will illustrate this by two examples: square lattice tilted along a principle lattice direction, and diagonally tilted decorated square. We leave a 
more complete discussion for future study.

\subsection{Square lattice tilted along a principal lattice direction}

This case was considered in Ref.~\onlinecite{SachdevMI}. However, the results there apply only if the resonant subspace
is limited by a large $|U_{3}|$. For small $|U_{3}|$ the physics is quite different, as we now describe.

\begin{figure}[h]
    \begin{center}
        \includegraphics[width=.5\textwidth]{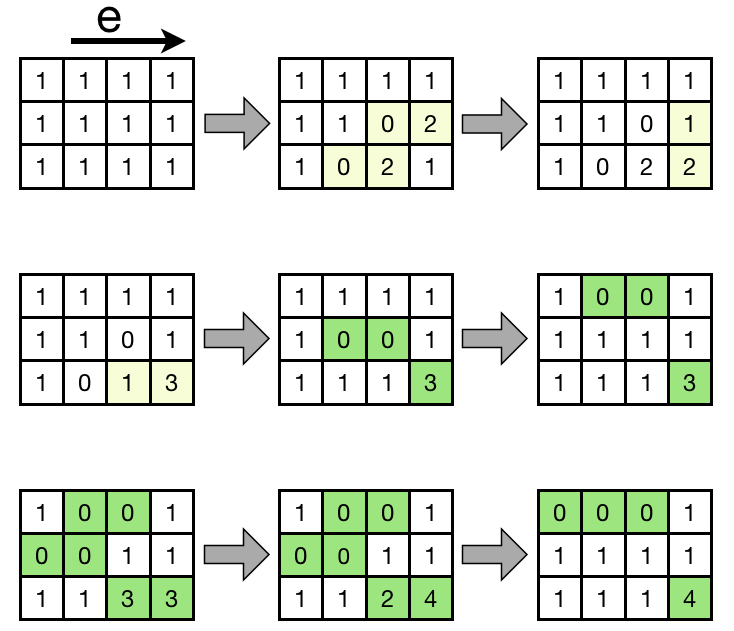}
    \end{center}
    \caption{(Color online) 
    Square lattice tilted along a lattice direction for $U_{3}=0$. The number in each square represents the number of bosons at that site. Here the occupancy of the parent Mott insulator is $n_0=1$. We illustrate resonant processes which eventually lead to the creation of lattice sites occupied by $4$ bosons.
    It is possible to continue this pattern, thus we obtain any arbitrarily large number $n_B$ of bosons on a single site, through resonant processes. In the end, only a total number of $n_B$ lattice sites remain changed (shown in green). Having $n_B$ bosons on a site reduces potential energy $n_B (n_B-1) \Delta/2$. One could fill the lattice with such patterns. Thus the energy of states resonantly connected to the parent Mott insulator is not bounded from below.
    }
    \label{fig:tetris}
 \end{figure}

In particular, we can no longer use equilibrium methods, as the energy of the subspace resonantly connected to the parent Mott insulator is not bounded from below, when $\lambda\rightarrow-\infty$.
%
 Once a particle-hole pair is created, the particle and hole are each free to move in the direction transverse to the tilt. It is then possible that two quasiparticles are adjacent to each other, in such a way that they can undergo a resonant transition. The system can resonantly reach states with an arbitrarily large number of bosons on a lattice site, as illustrated in \fref{fig:tetris}, leading to an arbitrarily large negative energy.

\subsection{Diagonally tilted decorated square lattice}
This lattice remains stable: the energy of the enlarged resonant subspace is bounded from below. This stability is related to the lower connectivity of the lattice structure, and the suppression of transverse superfluidity in \secref{sec:dsquare}. Let us consider the phases in the limit
$\lambda\rightarrow -\infty$.
We distinguish two cases: filling factor $n_0=1$, and filling factor $n_0>1$.
\subsubsection{Filling of parent Mott insulator: $n_0=1$}

\begin{figure}[h]
    \begin{center}
        \includegraphics[width=.5\textwidth]{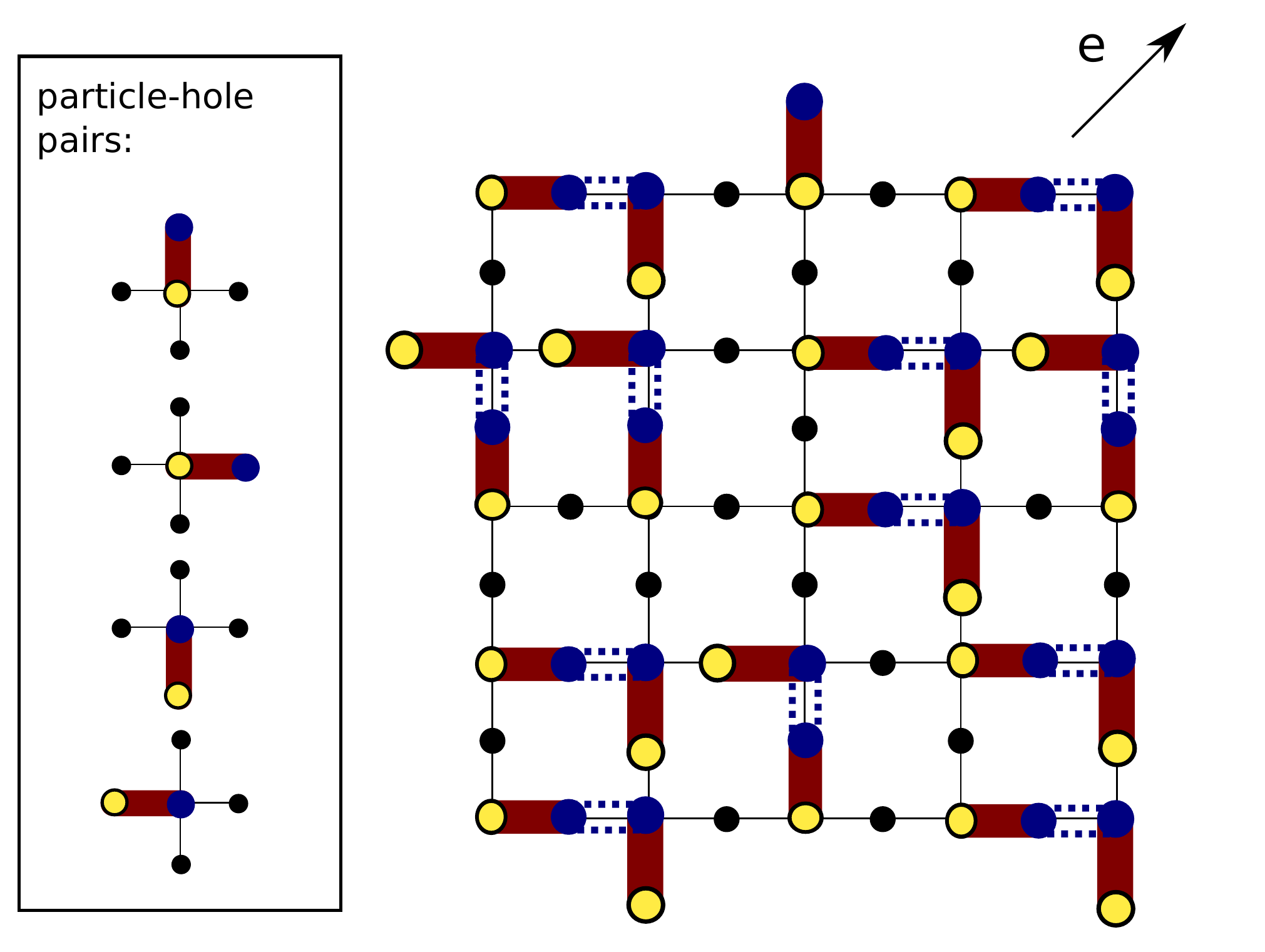}
    \end{center}
    \caption{(Color online) 
    Decorated square lattice in diagonal tilt, filling factor $n_0=1$, $U_3=0$.
    Quasiparticles are shown as dark blue circles, quasiholes as circles with yellow filling.
    Two neighboring quasiparticles can form another dipole bond $(2,2)\rightarrow(1,3)$ which reduces the energy by $\Delta$. This is indicated in the figure by dotted blue lines. Quasiholes contain no bosons, and they cannot form another dipole bond.
    To minimize its energy, the system not only
    maximizes the number of dipoles on the lattice, but also maximize the number of neighboring quasiparticles. The set of states fulfilling these requirements maps to the dimer coverings of a square lattice, and one such dimer covering is shown above.
    }
    \label{fig:dimer2}
 \end{figure}

Particle-hole pairs each reduce potential energy by $\Delta$, and they may be arranged in such a way that each quasiparticle is adjacent to another quasiparticle.
This allows for a third resonant transition, $(2,2)\rightarrow(1,3)$, which again reduces potential energy by $\Delta$.
As illustrated in \fref{fig:dimer2}, this always leads to a central site being occupied by three bosons. Two such sites are never adjacent to each other, so that an occupation number of four or more bosons on any site cannot be reached resonantly. The energy remains bounded from below.
In the limit $\lambda\rightarrow-\infty$ the resonantly connected subspace has a large number of ground states, and there is a one-to-one correspondence between this degenerate ground state manifold and the set of dimer packings on the square lattice.
%
A plaquette-flip term of the dimers appears in order $ 1 /{\lambda^{12}}$.
Thus we obtain a remarkable mapping of the effective Hamiltonian to the quantum dimer model.
Unlike the large $U_{3}$ case in \secref{decorated_lattices}, here the dimer model appears already for the singly-decorated square lattice, and so should be easier to realize experimentally.

\subsubsection{Filling of parent Mott insulator: $n_0>1$}

\begin{figure}[tb]
    \begin{center}
        \includegraphics[width=.5\textwidth]{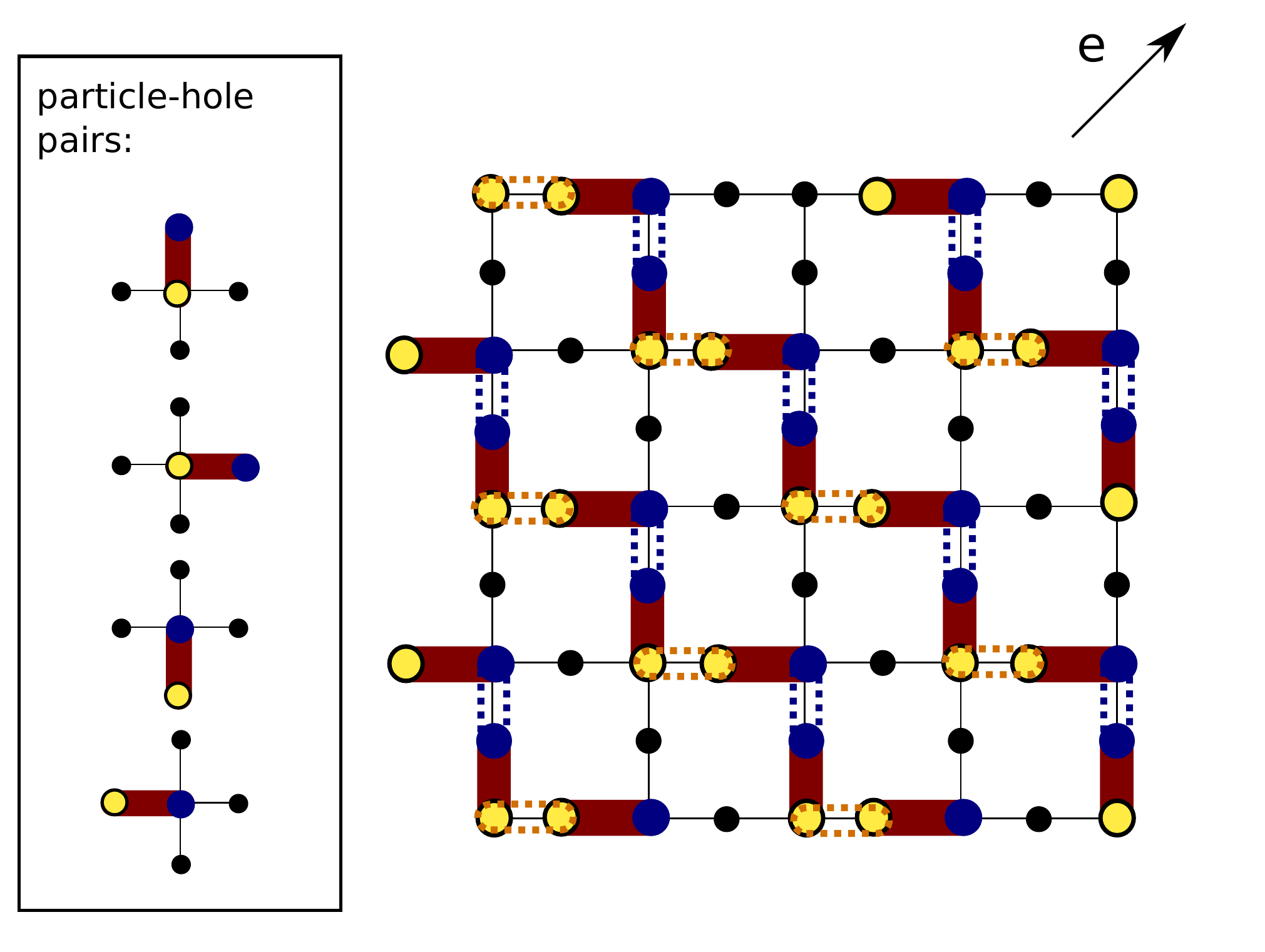}
    \end{center}
    \caption{(Color online) 
    Decorated square lattice in diagonal tilt, filling factor $n_0>1$, $U_3=0$.
    In this case `quasiholes' can also undergo a resonant transition, which is shown as orange dotted lines.
    There are two degenerate ground states with density-wave order: the one shown in the figure, and its symmetry related partner.
    }
    \label{fig:dimer3}
 \end{figure}

If the filling factor is two or larger, then two adjacent `quasiholes' can undergo a resonant transition $(n_0-1, n_0-1)\rightarrow(n_0-2, n_0)$, reducing the potential energy by $\Delta$.  
Potential energy is minimized if each `quasihole' and each `quasiparticle' can undergo such a transition, as illustrated in \fref{fig:dimer3}. This leads to a two fold degenerate ground state with density-wave order: all central sites are occupied by either $(n_0+2)$ or $(n_0-2)$ bosons, while all other sites contain $n_0$ bosons.

\section{Conclusions}
\label{sec:conc}

This paper has shown that there are rich possibilities for generating non-trivial quantum states
upon tilting a Mott insulator of bosons. Our classical intuition tells us that applying a strong tilt to a Mott insulator
should lead to a runaway instability of particles flowing downhill. However, quantum mechanically, for a single band model
in the absence of external
dissipation, and with a significant three-body interaction $U_{3}$, this does not happen. Instead particle motion
is localized in the direction of the tilt, and free motion is only possible in orthogonal directions. It is useful to introduce the idea of a resonant subspace of states which are strongly coupled to the parent Mott insulator when the tilt approaches its critical value.
Within this resonant subspace, we can define an effective Hamiltonian whose energy is bounded from below,
and which has well defined equilibrium quantum phases and phase transitions. We presented a variety of such
effective models here, and described general aspects of their phase diagrams. As in the previous work \cite{SachdevMI},
we found phases with Ising density wave order along the tilt direction, and superfluid/Luttinger liquid behavior in the direction
transverse to the tilt. More interestingly, we showed that on a variety of frustrated or decorated lattices, particle motion
can be suppressed also in directions transverse to the tilt. Then, the tilted lattices map onto quantum clock or dimer models,
with novel quantum liquid and solid phases.

We also briefly noted, in \secref{sec:tetris}, the situation when $\abs{U_{3}}$ was small. In some cases, the energy of the
resonant subspace is unbounded from below, and so non-equilibrium methods will be necessary to understand the physics.
However, for the simple case of the decorated square lattice, we found a mapping of the effective Hamiltonian to that
of the quantum dimer model in a limiting case, as illustrated in Fig.~\ref{fig:dimer2}.

It is clear that the models studied here have extensions to numerous other lattices and decorations, and that there
are many promising avenues for obtaining exotic phases. It would also be interesting to study the analogous models
of fermions, which could display metallic states associated with motion in the transverse directions.
In principle, all these models should be readily accessible
in experiments on trapped ultracold atoms.

\acknowledgements

We thank Waseem Bakr, Markus Greiner, Liza Huijse, and Jonathan Simon for many stimulating discussions.
We would like to thank Fabian Essler for pointing out that the decoupled one dimensional systems in
\secref{sec:decoupled} are Luttinger liquids only for $\lambda$ strictly~$-\infty$.
This research was supported by the National Science Foundation under grant DMR-0757145, by a MURI grant from AFOSR,
NSF grant DMR-0705472,
DMR-090647, AFOSR Quantum Simulation MURI, AFOSR MURI on
Ultracold Molecules, DARPA OLE program and Harvard-MIT CUA.
S.P. acknowledges support from the Studien\-stiftung des Deutschen Volkes.

\appendix

\section{Decorated square lattice in the $\lambda\rightarrow-\infty$ limit}

\subsection{Ground state correlation functions}

To calculate correlation functions in the exact ground state of
the effective clock model, Eq. (\ref{eq:exactgs}), we make use of
the following ``transfer matrix'' method. For an equal-amplitude
ground state, correlations of operators, $\hat O$, which are
diagonal in the basis of dipole coverings, can be related to a
corresponding classical problem
\begin{equation}
\ave{\hat O}=\frac{\bra{\psi_c}\hat O\ket{\psi_c}}{\braket{\psi_c}{\psi_c}}=\frac{\sum_M \bra{M}\hat O \ket{M}}{D_c}.
\end{equation}
Here, the sum over $M$ runs over all states that maximize the
number of dipoles and respect constraints \eqnref{dimerhardcore}
and \eqref{dimersconstraint}. We assume that all the states $\ket
M$ are properly normalized, $\braket{M}{M}=1$, and define
$D_c=\sum_M \braket{M}{M}=\dim(H_c)$ as the dimension of the
Hilbert space with constraint. The density operator $\hat n_{l,m}$
measuring the relative density at central site $(l,m)$ is diagonal
in the basis of dipole coverings; it has eigenvalues $\pm 1$.
Therefore, $\ave{\hat n_{(0,0)}\hat n_{(l,m)}}$ can be written as
\begin{equation}
\ave{\hat n_{(0,0)}\hat n_{(l,m)}}=\frac{N_+-N_-}{D_c},
\label{eqn:Acorr}
\end{equation}
where $N_\pm$ is the number of configurations which have $\bra{M}
\hat n_{(0,0)}\hat n_{(l,m)}\ket{M}=\pm1$. The problem of
calculating correlation functions of the ground state reduces to
counting the number of classically allowed dipole coverings. This
counting can be done using row-to-row transfer matrices.

Let $T$ be the transfer matrix for a row of length $N_x$ unit cells.
$T$ is a $4^{N_x}\times 4^{N_x}$ matrix and acts on the space of dipole configurations within that row.
Note that this Hilbert space includes configurations which are not
allowed by the constraint. Such configurations are excluded by
setting the corresponding matrix elements of $T$ to zero. The
matrix elements of $T$ are either $1$ or $0$, so that
$T_{c^\prime,c}=1$ if the sequence of two row configurations $c$,
$c^\prime$ contains no violations of the constraint, and
$T_{c^\prime,c}=0$ otherwise. The dimension of the Hilbert space
for a $N_x \times N_y$ lattice with periodic boundary conditions
is then given by the trace of powers of the transfer matrix,
\begin{equation}
D_c=\rm{Tr}({T^{N_y}}). \label{eq:d_c}
\end{equation}
A naive numerical implementation of Eq. (\ref{eq:d_c}) (without
using the sparseness of $T$) gives a complexity of $O(N_y
4^{3N_x})$, compared to going over all the states in the Hilbert
space with complexity $O(4^{N_x N_y})$. The values for $N_\pm$ can
be similarly found by including projection matrices in the trace,
for example
\begin{equation}
N_+={\rm Tr}\left(T^{N_y-m} P_{p;l} T^m P_{p;0}\right)+{\rm Tr}\left(T^{N_y-m} P_{h;l} T^m P_{h;0}\right).
\label{eq:ANpm}
\end{equation}
Here $P_{p;l}$ ($P_{h;l}$) is a projection matrix that projects
onto configurations which have a particle (a hole) at site
$l=0,1,\dots,N_x-1$ of a given row, respectively. Using this
method different correlations for the dipole-directions can also
be calculated.

\subsection{Connecting the ground state to a site-factorizable state}
\begin{figure}
    \includegraphics[width=.5\textwidth]{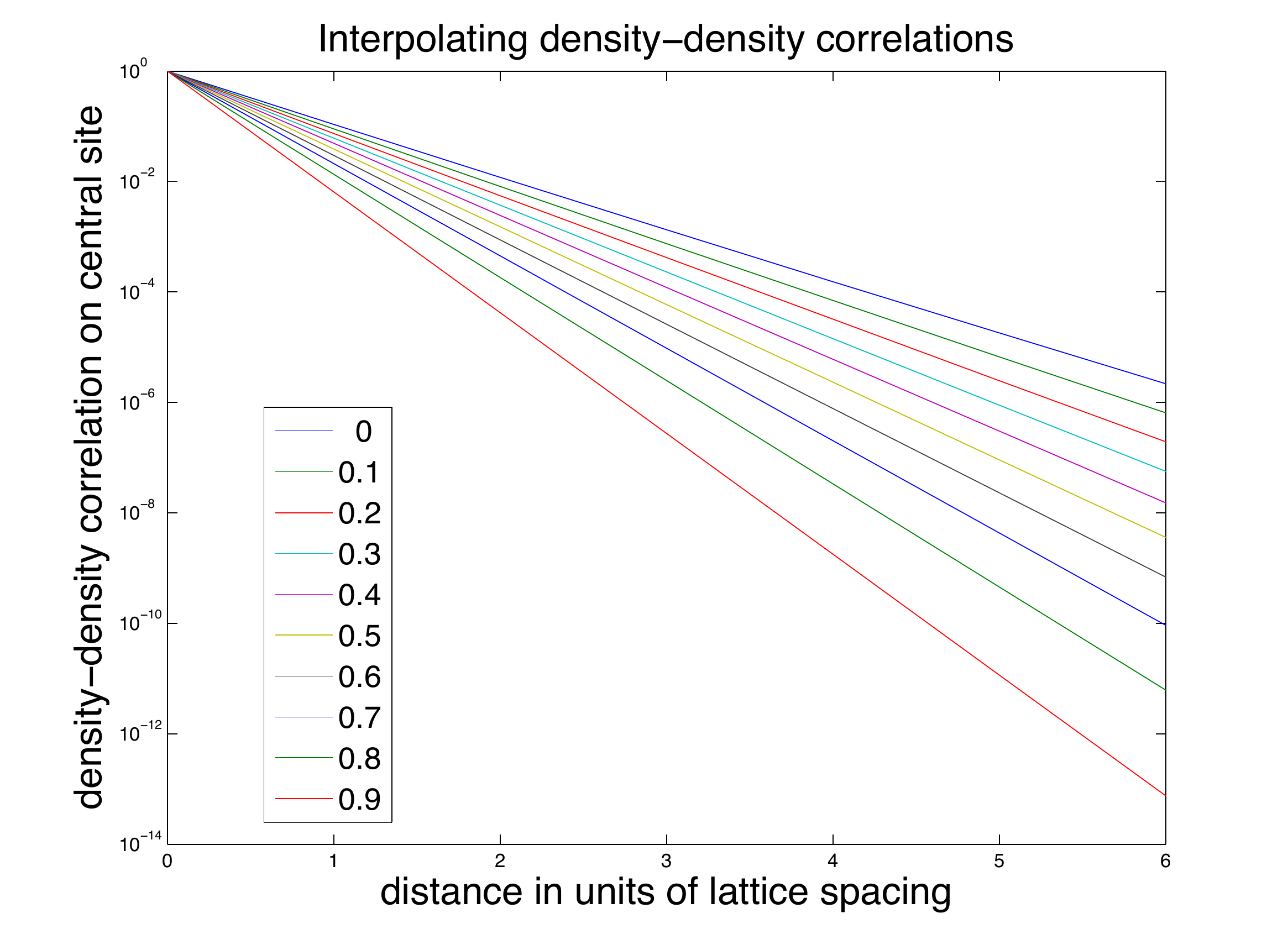}
\caption{(Color online) Density-density correlations of a wavefunction which interpolates between the ground states of $H^c_{\rm clock}$ and $H_{\rm clock}$, see text. Different colors stand for different values of the parameter~$\epsilon$. The correlation length does not diverge, it decreases as $\epsilon\rightarrow 0$. This provides evidence that the state $\ket{\psi_c}$ is topologically trivial. }
\label{fig:interpolating_constraints}
\end{figure}

In this appendix, we provide evidence that the ground state of
$H^c_{\rm clock}$, \eqnref{eq:exactgs}, is not topologically ordered. To show this,
we demonstrate that the ground state of $H^c_{\rm clock}$ can be smoothly
connected to the ground state of  $H_{\rm clock}$ (without the ``hard core''
constraint of the dipole-clocks). The ground state of  $H_{\rm clock}$ is a site
factorizable state, and therefore it is topologically trivial. The
fact that $\ket{\psi_c}$ can be connected to it, while keeping the
correlation length finite, is a strong indication that $\ket{\psi_c}$ does
not have topological order either.
To connect the ground state of $H^c_{\rm clock}$ to the ground state of $H_{\rm clock}$, \eqnref{eq:liquidhamiltonian}, we define the wave function
\begin{equation}
\ket{\psi(\epsilon)}=\sum_C \epsilon^{n(C)/2}\ket{C}.
\end{equation}
Here the sum over $C$ runs over all $4^N$ dipole coverings with one dipole per unit cell, including those which do not respect the constraint. The function $n(\epsilon)$ is an integer counting the number of colliding arrows in configuration $C$. All allowed configurations $M$ have $n(M)=0$. Thus changing the value of $\epsilon$ interpolates between the ground state of $H^c_{\rm clock}$,  $\ket{\psi_c}$, and the ground state of $H_{\rm clock}$,  $\ket{\psi_0}$,
\begin{eqnarray}
\ket{\psi(\epsilon=0)}=\ket{\psi_c},\\
\ket{\psi(\epsilon=1)}=\ket{\psi_0}.
\end{eqnarray}
As the density operator $\hat n_{(l,m)}$ is still diagonal in the basis of $C$, we can again use (generalized) transfer matrices to compute density-density correlation functions for any value of $\epsilon$. The result is shown in \fref{fig:interpolating_constraints}, and we find no divergence in the correlation length. This supports our conjecture that the ground state $\ket{\psi_c}$ is topologically trivial. Of course, we have not calculated \emph{all} possible correlation functions, so this is an evidence, but not a proof.

To calculate the correlation functions, we make use of equations \eqref{eqn:Acorr}, \eqref{eq:d_c}, and \eqref{eq:ANpm}, with one modification: the transfer matrix
$T(\epsilon)$ now depends on the parameter $\epsilon$ in the following way
\begin{equation}
(T(\epsilon))_{c^\prime,c}=\epsilon^{n(c^\prime,c)},
\end{equation}where $n(c^\prime,c)$
is the number of collisions in row-configuration $c^\prime$, plus the number of collisions {\emph{between}} rows $c$ and $c^\prime$.  The number of collisions in the other row, $c$, does not come in here, to avoid double counting.

\begin{thebibliography}{}
\bibitem{comment-symmetries} It is worth noting here that these effective Hamiltonians always
have full translational symmetry both parallel and transverse to
the tilt directions, unlike Eq.~(\ref{hubbard}) which is only
invariant under translations transverse to the tilt. Of course,
there is no equivalence between the physics parallel and
transverse to the tilt direction, and the lattice rotation
symmetry is reduced by the tilt, even in the effective
Hamiltonian.
\end{thebibliography}
\end{document}